\documentclass[pdflatex,sn-basic]{sn-jnl}

\usepackage{amsmath,amssymb}
\usepackage{bm}
\usepackage{graphicx}
\usepackage{booktabs}
\usepackage{array}
\usepackage{textcomp}
\usepackage{xcolor}
\usepackage{manyfoot}
\usepackage[section]{placeins}

\begin{document}

\title[Conservative DBS Formulation for PHREEQC-Coupled Reactive Transport]{A Conservative Darcy-Brinkman-Stokes Formulation for PHREEQC-Coupled Reactive Transport with Cell-Based Surface Complexation}

\author*[1,2]{\fnm{Hannah P.} \sur{Menke}}\email{H.Menke@hw.ac.uk}

\author[1,2]{\fnm{Florian} \sur{Doster}}

\author[1,2]{\fnm{Julien} \sur{Maes}}

\affil*[1]{\orgdiv{Institute for GeoEnergy Engineering}, \orgname{Heriot-Watt University}, \orgaddress{\city{Edinburgh}, \country{United Kingdom}}}

\affil[2]{\orgdiv{SET-IT (Subsurface Energy Transition \& Innovation Centre)}, \orgname{Heriot-Watt University}, \orgaddress{\city{Edinburgh}, \country{United Kingdom}}}

\abstract{We present a mass-conservative Darcy-Brinkman-Stokes (DBS)
formulation for reactive transport with cell-based surface complexation in
image-based porous media, coupled to PHREEQC and implemented in GeoChemFoam,
an OpenFOAM-based toolbox. Dissolved species are transported in the physical
mobile-water fraction, adsorbed species are stored as conservative cell-scale
surface inventories, and each reactive cell is coupled to PHREEQC. Reactive
surface area is prescribed independently as a cell field, so that internal
area left unresolved by segmentation of microporous phases can be represented
while DBS flow and transport are retained in the resolved pore space.

The implementation is verified with a test suite that isolates the coupling
of transport and chemistry from the construction of reactive surface area.
Aqueous and adsorbed mass balance is evaluated against a one-dimensional
analytical retardation benchmark and closed-system batch calculations. The
smooth-interface surface-area algorithms are tested on idealised two- and
three-dimensional geometries. In a two-scale lithium micromodel, DBS
reproduces fully resolved aqueous and sorbed inventories within $0.6\%$
across $Pe=0.1$--10.

The applications comprise Pb$^{2+}$ retardation by hydrous ferric oxide
coatings in a two-dimensional fracture network, competitive Mn$^{2+}$/Mg$^{2+}$
exchange in a three-dimensional Kelvin-cell sorbent, and Sr$^{2+}$/Ca$^{2+}$
exchange in Estaillades limestone. In the fracture-network and limestone cases,
the formulation yields spatially resolved aqueous and adsorbed inventories even
where reactive surface area is only partly resolved at the image scale. The
framework therefore supports conservative surface-complexation modelling in
both resolved and unresolved pore structures, with relevance to contaminant
retention, selective ion capture, and radionuclide transport.}


\maketitle

\noindent\textbf{Article Highlights}
\begin{itemize}
\item A cell-based method links pore-scale flow, solute transport, and surface chemistry.
\item Dissolved and surface-bound species enter the same local mass balance.
\item One surface-area field combines visible interfaces with unresolved internal area.
\end{itemize}

\section{Introduction}

Chemical exchange between the flowing aqueous phase and the solid matrix
governs the transport of dissolved species in porous media alongside advection
and diffusion. In many subsurface systems this exchange proceeds by surface
complexation, ion adsorption, and fast local equilibrium reactions, which
retard dissolved species, reshape concentration fronts, and alter the apparent
accessibility of pore space \citep{dzombak1990surface}. The effects are
strongest in multi-scale rocks that combine open macropores with microporous
matrix regions, where a single image may contain fast transport pathways, slow
exchange domains, and a large internal reactive area that the image cannot
resolve \citep{menke2021upscaling}. The same physics governs contaminant
retention and acid mine drainage treatment
\citep{webster1998trace,blowes2000treatment}, critical-mineral recovery and
direct lithium extraction \citep{reich2022lithium,vera2023environmental}, and
radionuclide transport for geological disposal
\citep{miller2010upscaling,dewindt2019reactive,bildstein2019rtm}, in all of
which selective adsorption or ion exchange may be controlled by surface area
that is not resolved at the image scale.

Three classes of models are available, and each is limited in a different way.
Classical Darcy-scale reactive transport models represent adsorption and
surface complexation through effective source terms, bulk sorption
coefficients, or equilibrium closures applied to the continuum volume
\citep{steefel1990kinetic,steefel1994coupled,steefel2015reactive,mayer2002multicomponent,xu2006toughreact,lichtner1985continuum}.
They remain indispensable for large-domain prediction, but they require
constitutive upscaling assumptions for both transport and reactive surface
area, and they cannot reproduce the spatial heterogeneity of surface inventory
that image-based data make visible \citep{battiato2011applicability}. Fully
resolved pore-scale models, including direct numerical simulation
\citep{molins2021simulation,molins2015reactive,molins2012investigation,starchenko2018development},
lattice Boltzmann methods \citep{yoon2015lattice,kang2007lattice}, and
explicit-boundary DBS approaches
\citep{maes2018direct,maes2021geochemfoam,soulaine2017mineral}, instead locate
reactive uptake at the meshed grain surface. This is appropriate when the
dominant surface area coincides with the resolved interface between pore and
grain, but it is bounded by image resolution: once internal micro-scale area
falls below the mesh scale, a boundary-resolved description is no longer
faithful to the physical geometry, and the cost of resolving it rises steeply.
Pore-network and other reduced pore-scale models occupy a middle ground
\citep{blunt2001flow}, but they depend on idealised throat-and-pore
representations and do not deliver the voxel-by-voxel field description used in
image-based DBS simulations.

The consequences are specific in image-based simulations of carbonate and other
multi-scale porous media \citep{menke2015dynamic, menke2018fourd,
menke2021upscaling,wildenschild2013xray,blunt2013porescale}. Segmented images
routinely separate macropore, microporous, and nearly solid regions, but the
reactive internal surface of the microporous phase is not the same object as
the explicitly meshed boundary of the coarse computational cells. Ignoring that
internal area under-predicts reactive uptake. Inserting it as an ad hoc
volumetric sink leaves both the conservation properties and the meaning of the
cell-scale surface inventory unclear.

PHREEQC and its reaction-module form, PhreeqcRM, already provide the required
chemistry: aqueous speciation, surface complexation, equilibrium and kinetic
reactions, and operator-split coupling to transport simulators
\citep{parkhurst2013phreeqc, parkhurst2015phreeqcrm,charlton2011modules}.
Existing PHREEQC-coupled workflows serve structured-grid continuum models and
operator-split multi-component solvers well \citep{jyoti2021validation}. The
open question is how to embed that engine in a transport formulation consistent
with unresolved image-based geometry. The closest existing approach is
\texttt{porousMedia4Foam} \citep{soulaine2021porousmedia4foam,
pavuluri2022reactive}, which couples a micro-continuum OpenFOAM solver to
PHREEQC at pore, hybrid, and Darcy scales, but which applies chemistry through
kinetic or equilibrium flux conditions at the diffuse fluid-solid interface. It
therefore inherits the limitation of the fully resolved methods, in that the
reactive surface must overlap the resolved computational boundary, and it draws
no explicit distinction between the numerical DBS porosity used to regularize
transport and the physical mobile-water fraction that governs chemistry.
Earlier PHREEQC-coupled cases within GeoChemFoam itself
\citep{maes2022heat,maes2022vos} share this restriction: they support reactive
transport and surface complexation where the reactive surface coincides with an
explicit mesh boundary, which remains valid but does not cover the case in
which the dominant reactive area lies inside the coarse cell.

In this work we develop a DBS-compatible reactive transport formulation in
which the reactive surface is a property of the cell rather than of the mesh
boundary. Aqueous transport is written for the physical mobile-water fraction;
unresolved reactive surface area is prescribed as a cell field and carried as a
cell-centred adsorbed inventory; local PHREEQC chemistry transfers mass
conservatively between the aqueous and adsorbed states; and no heuristic
redistribution of source terms is used. Two consequences follow. The numerical
DBS porosity and physical mobile-water fraction have distinct roles: the former
sets flow resistance \citep{soulaine2016micro}, whereas the latter sets aqueous
storage and chemical solution volume. The saturated calculations reported here
use the same numerical field for both quantities. The surface inventory is
defined separately at cell centres and driven by an unresolved internal
surface-area density, rather than reconstructed on mesh
faces. Together these choices let the method act on microporous matrix regions
instead of only on explicitly resolved boundaries between pore and grain. The
governing equations are posed on general finite-volume meshes; the image-based
cases used here are the motivating application.

The solver that results, \texttt{reactiveTransportDBSFoam}, couples
conservative DBS transport to PHREEQC-based surface complexation at the cell
scale. A one-dimensional retardation benchmark compares the coupled advection,
diffusion, and sorption response with a boundary-matched finite-domain
reference, while closed-system batch calculations calibrate the dilute surface
response and test the aqueous and adsorbed mass balance.
The surface-area algorithms are verified on controlled geometries of known area
and on image-based micromodels, and a resolved-to-DBS two-scale benchmark of
selective Li$^{+}$ uptake tests whether the upscaled cell-based description
reproduces a fully resolved calculation of the same system. The formulation is
then applied to three image-based geometries of increasing complexity:
Pb$^{2+}$ uptake on hydrous ferric oxide (HFO) in a two-dimensional fracture
network, as an analogue for heavy-metal remediation in fractured aquifers and
permeable reactive barriers; selective Mn$^{2+}$ uptake from a Mg$^{2+}$-rich
brine on HFO in a three-dimensional Kelvin-cell sorbent, as an analogue for
critical-metal capture; and Sr$^{2+}$ migration in Estaillades limestone, as an
analogue for radionuclide transport in a deep geological repository (DGR).
Together these cases determine how resolved pore geometry and unresolved
internal surface area interact to control the spatial distribution of aqueous
and adsorbed species in multi-scale porous media.
\section{Model Description}
\label{sec:model}

\subsection{Field Definitions}

The DBS micro-continuum formulation \citep{brinkman1949calculation,soulaine2016micro,
soulaine2024microcontinuum} represents the pore medium as a continuum in which
a cell-centred porosity field $\epsilon$ varies smoothly from
zero (solid) to unity (open pore). The steady incompressible flow is governed
by the Darcy-Brinkman equations:
\begin{align}
\nabla \cdot \bm{u}_D &= 0,
\label{eq:continuity} \\
-\nabla p + \nabla \cdot \bigl(\mu_{\mathrm{eff}} \nabla \bm{u}\bigr)
- \frac{\mu}{K(\epsilon)}\,\bm{u}_D &= \bm{0},
\label{eq:brinkman}
\end{align}
where $\bm{u}_D = \epsilon\bm{u}$ is the Darcy velocity,
$\mu$ is the fluid viscosity, $\mu_{\mathrm{eff}}$ is an effective
viscosity that smoothly recovers the Stokes equation in open-pore cells
\citep{soulaine2016micro}, and $K(\epsilon)$ is a permeability
function. The drag coefficient $\mu/K(\epsilon)$ therefore acts
as a large resistance term in solid-like cells and vanishes in open-pore cells,
so Eq.~\eqref{eq:brinkman} reduces to the Stokes equation.

For an image-derived cell, let $f_k$ be the geometric fraction of source voxels
assigned to segmented phase $k$, with $\sum_k f_k=1$, and let $\phi_k$ be the
end-member porosity assigned to that phase. The cell DBS porosity is
\begin{equation}
\epsilon=\sum_k f_k\phi_k.
\label{eq:phase_porosity}
\end{equation}
The fraction $f_k$ describes phase occupancy, whereas $\phi_k$ specifies the
water-filled fraction of that phase. A cell wholly occupied by a microporous
phase therefore has $f_{\mathrm{micro}}=1$ and
$\epsilon=\phi_{\mathrm{micro}}<1$. Nominal solid is assigned a small non-zero
porosity as a numerical DBS floor.

The mobile-water fraction $\theta$ is the mobile-water volume per bulk cell
volume. It governs aqueous storage, diffusion, and PHREEQC solution volume,
whereas $\epsilon$ governs DBS flow resistance and relates intrinsic velocity
to Darcy velocity. The saturated calculations reported here set
$\theta=\epsilon$ in chemically active cells. Cells with
$\theta\leq\theta_{\mathrm{mask}}$ are excluded from chemistry.

Reactive surface area is specified independently as the cell-centred bulk-volume
density $a_{\mathrm{surf}}$ (m$^2$\,m$^{-3}_{\mathrm{bulk}}$). All aqueous
species, $\theta$, $a_{\mathrm{surf}}$, and adsorbed surface fields are
cell-centred; transport fluxes are evaluated on cell faces.

\subsection{Reactive Surface Area Field}
\label{sec:aSurf}

Reactive area is present only for selected interfaces and phases assigned a
non-zero intrinsic density. Its resolved and unresolved contributions are
carried in the fixed cell-centred field $a_{\mathrm{surf}}$ (units:
m$^2\,\mathrm{m}^{-3}_{\mathrm{bulk}}$), the sum of two terms:
\begin{equation}
a_{\mathrm{surf}} = a_{\mathrm{surf}}^{\mathrm{interface}} +
a_{\mathrm{surf}}^{\mathrm{intrinsic}}.
\label{eq:asurf}
\end{equation}

The interface term is reconstructed for an ordered pair of segmented labels.
One label is the owner and the second is the other phase; their union restricts
the extracted interface, and the owner identifies the side used for area
assignment. In \texttt{occupancyGaussian}, the Gaussian-filtered owner
indicator is normalised by the Gaussian-filtered pair mask, and the $0.5$
level is extracted.

Three-dimensional triangle area is divided by bulk cell volume to obtain
$a_{\mathrm{surf}}^{\mathrm{interface}}$. In quasi-two-dimensional images,
contour length is multiplied by physical out-of-plane thickness before
division by cell volume. For quasi-two-dimensional cell-field construction,
\texttt{occupancyGaussian} uses weighted redistribution over owner-side
first-shell contact cells. Pieces with empty support contribute zero area
to the deposited field. Detailed support and partition rules are given in
Online Resource~1.

\begin{table*}[htbp]
    \centering
    \caption{Interface reconstruction parameters. All cases use
    \texttt{occupancyGaussian} with an isovalue of $0.5$.
    The ordered pair is owner $\rightarrow$ other; the owner defines the
    indicator used for reconstruction. In cell-field constructions,
    reconstructed area is assigned on the owner side.}
    \label{tab:surfaceConstruction}
    \scriptsize
    \setlength{\tabcolsep}{3pt}
    \begin{tabular}{@{}p{0.28\textwidth}p{0.12\textwidth}p{0.48\textwidth}@{}}
        \toprule
        Case & $\sigma$ (vox) & Pair (owner $\rightarrow$ other) \\
        \midrule
        Circular post & 1.50
            & exterior/pore 255 $\rightarrow$ post/solid 0 \\
        Sphere & 0.75
            & exterior/pore 255 $\rightarrow$ sphere/solid 0 \\
        HM12 solid-grain DBS & 0.75
            & pore 255 $\rightarrow$ solid 0 \\
        2D four-fracture & 0.75
            & micropore 1 $\rightarrow$ pore 2 \\
        3D Kelvin & 0.75
            & micropore 2 $\rightarrow$ pore 1 \\
        3D Estaillades & 0.74
            & pore 1 $\rightarrow$ solid 3 \\
        3D external area & 0.74
            & pore 1 $\rightarrow$ remapped solid 3 \\
        \bottomrule
    \end{tabular}
\end{table*}

The interface term captures only boundaries between grain and fluid that are
visible at the resolved image scale. It does not detect
reactive area that lies entirely \emph{within} a voxel, such as sub-voxel
surface roughness of a grain or the internal pore network of a microporous
phase. Those surfaces are below the image resolution by construction.

The second term, $a_{\mathrm{surf}}^{\mathrm{intrinsic}}$, accounts for this
sub-voxel contribution. An end-member surface density $s_k$
(units: m$^2$\,m$^{-3}_{k}$) is prescribed per unit volume of each segmented
phase $k$ (for example pore, micropore, or dense solid). It may represent
sub-voxel surface such as the
Brunauer-Emmett-Teller (BET) surface area of a microporous grain measured by
N$_2$ adsorption, or the surface roughness of a dense solid grain measured on
powdered material. Mass-specific measurements are converted to phase-volume
densities using the corresponding mass per unit phase volume.

The intrinsic contribution is assembled by weighting each phase density by
its geometric fraction $f_k$ within the computational cell:
\begin{equation}
a_{\mathrm{surf}}^{\mathrm{intrinsic}} = \sum_k f_k\,s_k.
\label{eq:asurf_intrinsic}
\end{equation}
Thus $f_k$ converts the phase-volume density $s_k$ to a bulk-volume cell
contribution. A density already expressed per bulk cell volume represents
that contribution directly. The segmented image supplies $f_k$, whereas
$s_k$ is prescribed from complementary measurements or finer-scale
characterisation.

$s_k$ may be non-zero for any label, including labels selected for interface
reconstruction. The same segmented label may contribute through both terms:
$a_{\mathrm{surf}}^{\mathrm{interface}}$ gives the image-scale outer surface
of the grain, and $a_{\mathrm{surf}}^{\mathrm{intrinsic}}$ adds the
sub-voxel surface area that the image resolution cannot capture. If
$s_k=0$ for all labels, the total reduces to the purely geometric image-scale
surface. Both contributions are pre-computed using the segmented image and
prescribed phase properties before the flow simulation begins and are held
constant throughout the reactive-transport calculation.

Both contributions enter the geochemical module through the same bulk-volume
field. The \texttt{phreeqcMixture} builds used for the reported calculations
supply surface area per representative bulk litre to PHREEQC as
\begin{equation}
a_{\mathrm{surf}}^{\mathrm{bulk,L}} = \frac{a_{\mathrm{surf}}}{10^{3}},
\label{eq:asurf_bulk_litre}
\end{equation}
For surface master $k$, the site density $N_k$ (mol\,m$^{-2}$) gives a
surface-master inventory per bulk volume of $N_k a_{\mathrm{surf}}$
(mol\,m$^{-3}_{\mathrm{bulk}}$). These conversions use the bulk-volume basis
without division by $\theta$.

Surface areas estimated from nano-computed tomography
(nano-CT), micro-computed tomography (micro-CT), focused ion beam scanning
electron microscopy (FIB-SEM), or other image data are first expressed on a
bulk-volume basis (m$^2$\,m$^{-3}_{\mathrm{bulk}}$), consistent with
Eqs.~\eqref{eq:asurf} and~\eqref{eq:asurf_intrinsic}. Image-derived surface areas
reported per microporous-phase volume or per grain mass are converted to this
bulk-volume basis before Eq.~\eqref{eq:asurf_bulk_litre} is applied.

The same fixed porosity field is used for DBS transport and for the PHREEQC
solution volume. Cells in
which $\theta$ falls below the chemistry mask threshold
$\theta_{\mathrm{mask}}$ are treated as chemically inactive and their surface
terms are suppressed.

\subsection{Conservative Aqueous Transport}

For each aqueous species with concentration $C_i$, the conservative DBS
transport equation is written as
\begin{equation}
\frac{\partial (\theta C_i)}{\partial t}
+ \nabla \cdot (\bm{u}_D C_i)
- \nabla \cdot (\theta D_i \nabla C_i) = 0,
\end{equation}
where $\bm{u}_D$ is the Darcy velocity satisfying
Eqs.~\eqref{eq:continuity} and~\eqref{eq:brinkman}, and $D_i$ denotes the
diffusivity used in the transport equation. In fully resolved macro-pore cells
($\theta \approx 1$), $D_i$ is the free molecular diffusivity. In cells that
represent microporous material ($0 < \theta < 1$), $D_i$ is interpreted as an
effective diffusivity, commonly $D_i = D_i^{\mathrm{mol}}/\tau$, where
$D_i^{\mathrm{mol}}$ is the molecular diffusivity and $\tau \geq 1$ is the
tortuosity factor of the micropore network. In either case, the $\theta$ factor
multiplying $D_i$ in the diffusion term accounts for the reduced water volume
available for transport. Reactions enter through the operator-split PHREEQC
step described below, not as a source term in this equation.

For a vector of aqueous species $\bm{C}$, the same balance is
\begin{equation}
\frac{\partial (\theta \bm{C})}{\partial t}
+ \nabla \cdot (\bm{u}_D \bm{C})
- \nabla \cdot (\theta \bm{D} \nabla \bm{C}) = \bm{0},
\end{equation}
where $\bm{D}$ collects the species diffusivities. This notation is used for
the split operators below.

\subsection{Cell-Based Surface Inventory}

For the cell-based unresolved surface carried by the DBS formulation, the
conservative local inventory is expressed
in the current solver field variables as
\begin{equation}
M = \theta C + a_{\mathrm{surf}}\Gamma,
\label{eq:inventory}
\end{equation}
where $\Gamma$ is the cell-centred surface-species concentration
(mol\,m$^{-2}$ of reactive surface). The first term is the aqueous amount per
bulk volume, the second the adsorbed amount per bulk volume. Porosity enters
the aqueous term once, and the bulk-based surface-area field carries no
porosity scaling.

For multiple surface species the local bulk inventory is the sum of the aqueous
and adsorbed contributions over species; single-species notation is used
below.

\subsection{Local Chemistry Coupling}

The chemical update is local to each cell. Over one cycle of transport and
chemistry, the transport step first advances the aqueous species in
conservative finite-volume form. The updated cell state, together with the mobile-water
fraction and cell-based surface state, is then mapped into PHREEQC
\citep{parkhurst2013phreeqc} via the PhreeqcRM interface
\citep{parkhurst2015phreeqcrm}. PHREEQC solves the local chemistry at
\emph{thermodynamic equilibrium}, treating surface complexation as
instantaneous relative to transport within each timestep. This holds when the
surface equilibration time is short compared with the time for advection and
diffusion across a cell. The same interface supports kinetic surface reactions,
which are not used here.

The solver uses first-order Lie splitting \citep{carrayrou2004operator}. The
splitting introduces a global error of $\mathcal{O}(\Delta t)$ in the
operator-split solution; the local error over one step is
$\mathcal{O}(\Delta t^2)$. For the timesteps used here, the advective
Courant-Friedrichs-Lewy (CFL) number remains $O(1)$, so the
operator-splitting error is of the same formal order as the first-order
transport discretisation. The resulting algorithm keeps the conservative
transport update and the local PHREEQC equilibrium update as distinct
operators within each timestep.

At the algorithmic level, one timestep of size $\Delta t$ is:
\begin{align}
\widetilde{\bm{C}}^{\,n+1} &= \mathcal{T}_{\Delta t}(\bm{C}^{n}, \theta, \bm{u}_D, \bm{D}),\\
(\bm{C}^{n+1}, \bm{\Gamma}^{n+1}) &=
\mathcal{R}_{\Delta t}(\widetilde{\bm{C}}^{\,n+1}, \bm{\Gamma}^{n}, \theta, a_{\mathrm{surf}}),
\end{align}
where $\mathcal{T}_{\Delta t}$ is the conservative finite-volume transport
operator (advancing $\theta\bm{C}$ in DBS form) and $\mathcal{R}_{\Delta t}$
is the local PHREEQC equilibrium operator that updates both aqueous and
surface states simultaneously within each cell.

\subsection{Analytical Dilute Limit}

In the dilute linear-equilibrium limit, surface complexation reduces to a
Henry-type isotherm. The standard Henry coefficient is $K_s = \Gamma / C$
(units: m), the ratio of surface concentration $\Gamma$ (mol\,m$^{-2}$) to
aqueous concentration $C$ (mol\,m$^{-3}$):
\begin{equation}
\Gamma = K_s C.
\end{equation}
$K_s$ is obtained from a pair of batch equilibrations at two aqueous
concentrations, $K_s = \Delta\Gamma / \Delta C$, which returns the coefficient
in the field units used by the retardation formula.

Substituting into the inventory Eq.~\eqref{eq:inventory} and taking the
time derivative yields the retarded 1D transport equation
\begin{equation}
\left(
\theta + a_{\mathrm{surf}}K_s
\right)\frac{\partial C}{\partial t}
+ \frac{\partial (u_D C)}{\partial x} = 0.
\end{equation}
The corresponding retardation factor is
\begin{equation}
R = 1 + \frac{a_{\mathrm{surf}}\,K_s}{\theta}.
\label{eq:retardation}
\end{equation}
Both $a_{\mathrm{surf}}$ and $\theta$ can vary spatially, so
Eq.~\eqref{eq:retardation} also gives the local retardation at any point in
a heterogeneous domain. The classical semi-infinite advection--diffusion
solution follows \citet{ogata1961solution}. For comparison with the finite 1D
calculation, the reference problem is instead evaluated on $[0,L]$ with the
same fixed-concentration inlet and zero-gradient outlet as the numerical case,
using the effective velocity $u_D/(\theta R)$ and diffusivity $D/R$.
\section{Numerical Implementation}

The implementation couples conservative Darcy-Brinkman-Stokes (DBS) transport
with cell-centred PHREEQC chemistry and cell-based surface complexation in a
single operator-split formulation.

\subsection{Field Locations and Discretisation}

All chemistry-bearing state is cell-centred. The solution species, $\theta$,
$a_{\mathrm{surf}}$, and the adsorbed surface-species fields are stored at cell
centres, as is the Darcy velocity $\bm{u}_D$. Transport uses face interpolation
to assemble conservative fluxes on cell faces, while surface chemistry is
evaluated as a cell-local update. The additive decomposition of
$a_{\mathrm{surf}}$ into interface and intrinsic contributions is defined in
Section~\ref{sec:aSurf}
(Eqs.~\eqref{eq:asurf} and~\eqref{eq:asurf_intrinsic}).

Time integration uses the first-order implicit Euler scheme, matching the
first-order Lie splitting of the transport and chemistry operators. The
convective term $\nabla\cdot(\bm{u}_D C_i)$ is discretised with the standard
first-order upwind scheme, which is bounded and monotone on the unstructured
image-based meshes. The diffusive term $\nabla\cdot(\theta D_i \nabla C_i)$
uses a second-order Gauss linear scheme. The Cartesian image cases run without
non-orthogonality correction, which their orthogonal grids do not require. The
combination is first-order accurate in both time and space for
advection-dominated transport.

\subsection{PHREEQC Coupling}

The reported calculations use \texttt{reactiveTransportDBSFoam} with
\texttt{mixtureType phreeqcMixture} in
\texttt{constant/thermoPhysicalProperties}. This mixture wraps the PhreeqcRM
interface \citep{parkhurst2015phreeqcrm}. At
initialisation, each active computational cell is registered as a separate
PHREEQC solution. When surface complexation is active, a PHREEQC \texttt{SURFACE}
block is also created for each reactive cell. For surface master $k$, its
capacity per representative bulk volume is
\begin{equation}
q_{\mathrm{sites},k} = N_k \, a_{\mathrm{surf}},
\label{eq:nsites}
\end{equation}
where $N_k$ (mol\,m$^{-2}$) is the user-prescribed site density for surface
master species $k$.
The OpenFOAM dictionary stores $N_k$ in kmol\,m$^{-2}$, following the native
OpenFOAM amount unit; it is therefore multiplied numerically by $10^3$ when
interpreted as mol\,m$^{-2}$. Each surface-master density is supplied as an
OpenFOAM field, allowing spatially heterogeneous site densities.

The total $a_{\mathrm{surf}}$ field is computed before the simulation begins
(Section~\ref{sec:aSurf}), written as an OpenFOAM volume field, and read
explicitly by \texttt{reactiveTransportDBSFoam}. It is held constant throughout
the reactive-transport calculation; it is not updated by the flow or chemistry
solvers. The area and site capacity are transferred when the per-cell
PHREEQC surface templates are created and remain internal to PhreeqcRM during
the time loop. Consequently, the present implementation supports spatially
heterogeneous but time-independent $\epsilon$ and $a_{\mathrm{surf}}$; changing
either field during reactive transport would require rebuilding the PHREEQC
surface state. The surface mode is optional and reduces exactly to
aqueous-only mode when $a_{\mathrm{surf}} = 0$ everywhere.
The interface algorithms, owner support, and case-specific parameters are given
in Section~\ref{sec:aSurf} and Table~\ref{tab:surfaceConstruction}. Both the 2D
line-length--thickness route and the 3D triangle-area route produce
$a_{\mathrm{surf}}$ in m$^2$\,m$^{-3}_{\mathrm{bulk}}$ for the same
cell-local PHREEQC mapping.

Cells with negligible mobile-water fraction
($\theta < \texttt{epsChemistryMin}$) are excluded from the fixed
finite-volume-to-PhreeqcRM mapping and do not contribute to chemistry.

The chemistry update is local to each mapped cell. Each solver process
constructs PhreeqcRM with one worker thread, and OpenFOAM domain decomposition
provides process-level parallelism.
The \texttt{runCells()} call advances the chemistry cells held by that process.

\subsection{Per-Timestep Algorithm}

At startup, \texttt{reactiveTransportDBSFoam} reads the fixed porosity,
velocity, pressure, aqueous-species, and total surface-area fields. The
\texttt{phreeqcMixture} constructor establishes a fixed mapping between active
finite-volume cells and PhreeqcRM cells and caches the surface-area and
site-capacity data. Unless checkpoint restoration is explicitly requested, the
chemistry state is initialised from the written aqueous and surface fields at
the selected start time. When \texttt{phreeqcRestoreCheckpoint} is enabled, a
complete matched checkpoint restores the serialised PhreeqcRM state;
restoration fails if either checkpoint component is absent or inconsistent.

At each timestep, the aqueous transport equations are advanced in conservative
DBS form using the implicit Euler upwind discretisation described above. A
single call to \texttt{species}\allowbreak\texttt{Mixture.}\allowbreak\texttt{correct()} transfers the updated aqueous
and surface fields to PhreeqcRM, refreshes the fixed porosity values, clips
negative aqueous inputs to zero subject to a fail-fast relative added-mass
limit, executes \texttt{runCells()}, and copies the equilibrated aqueous and
surface species back to the finite-volume fields. At configured output times,
checkpoint-enabled runs write the aqueous and surface fields together with the
corresponding rank-local PhreeqcRM checkpoint and publish its completion marker
only after the field write succeeds. Conservation is assessed from the written
fields using the domain-integrated bulk inventory
$M_{\mathrm{total}} = \sum_{\mathrm{cells}} V_{\mathrm{cell}}
\left( \theta C + a_{\mathrm{surf}}\Gamma \right)$,
where $V_{\mathrm{cell}}$ is the cell volume. In
a closed domain this quantity should remain constant to within discretisation
error; in a flow-through domain it should balance the net flux through the
inlet and outlet faces.

\begin{figure}[htbp]
    \centering
    \includegraphics[width=0.96\linewidth]{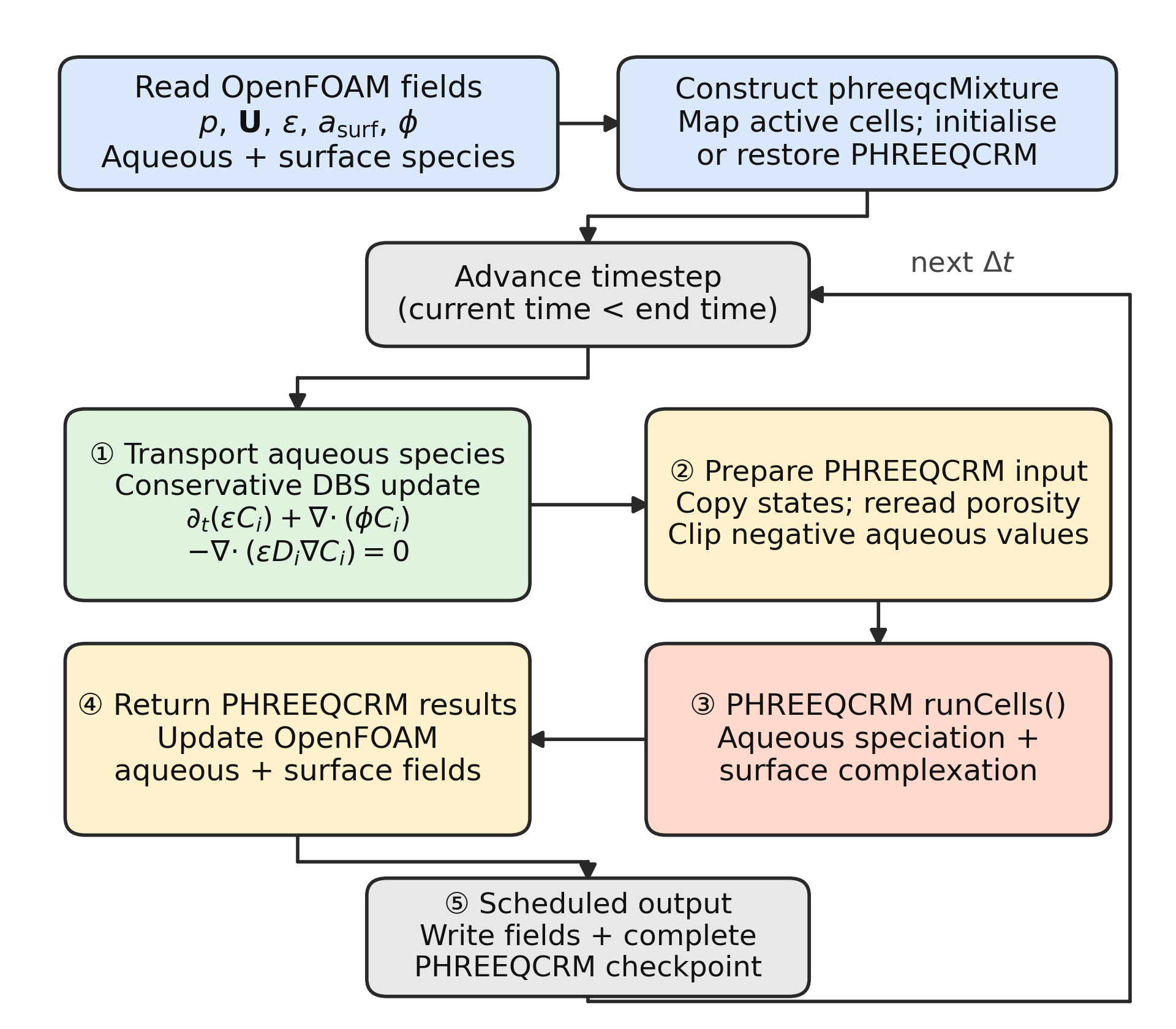}
    \caption{Source-traced solution procedure for
    \texttt{reactiveTransportDBSFoam} with \texttt{phreeqcMixture}. The
    constructor establishes the active-cell mapping and initial PhreeqcRM
    state. Each timestep advances all aqueous species in conservative DBS
    form, after which \texttt{speciesMixture.correct()} transfers the aqueous
    and surface state to PhreeqcRM, executes \texttt{runCells()}, and returns
    the updated species fields. At output times, it writes the fields and, when
    enabled, a matching complete PhreeqcRM checkpoint.}
    \label{fig:dbsWorkflow}
\end{figure}

Figure~\ref{fig:dbsWorkflow} illustrates the algorithm. Solver selection is
made through \texttt{mixtureType}: the surface-complexation cases use
\texttt{phreeqcMixture}, whereas the upstream kinetic-mineral formulation uses
\texttt{inertMultiComponentMixture}. The chemistry state remains cell-centred
throughout the loop: transport acts through the aqueous face fluxes, while
PHREEQC updates the cell-local aqueous and adsorbed inventories after each
transport solve. The unresolved surface reaction stays a cell-local update
rather than an equivalent face flux.
\section{Validation}\label{sec:validation}

The validation ladder separates four questions. The closed-batch and transient
1D benchmarks separately test conservative aqueous--surface state transfer and
coupled reactive transport independently of image geometry. Controlled
geometric tests then verify how the reactive surface-area field
$a_{\mathrm{surf}}$ is constructed from segmented images, and image-based
micromodel comparisons put that same construction to work in reactive
transport. A resolved-to-DBS multiscale micromodel benchmark closes the ladder
by testing whether the upscaled surface-area representation preserves
sorption-controlled transport when part of the image is treated as an
unresolved reactive matrix.

For the image-derived geometric benchmarks, \texttt{snappyHexMesh} provides the
resolved-wall reference: an established OpenFOAM route to a body-fitted mesh
from a segmented interface between pore and solid, on which reactive area and
surface inventory can be evaluated directly. Its practical range sets the range
of the comparison. Meshing becomes unreliable or fails for the segmented
volumes considered here once the image size reaches order $600^3$ voxels, and
the snapped mesh carries no unresolved DBS matrix volume or intrinsic sub-voxel
area as a cell-supported inventory. The DBS surface-area fields are therefore
compared against the resolved-wall area distribution and sorption response
wherever an explicit-wall reference can be generated.

For validation quantities compared without independent normalization, we
report the physical-measure-weighted relative $L_1$ discrepancy,
\begin{equation}
    E_1(q) = 100\,
    \frac{\sum_i w_i\lvert q_i-q_i^{\mathrm{ref}}\rvert}
         {\sum_i w_i\lvert q_i^{\mathrm{ref}}\rvert}.
    \label{eq:relativeL1}
\end{equation}
The weights $w_i$ represent the appropriate pore volume, bulk-cell volume,
interfacial area, or streamwise-bin width. When each $q_i$ is already an
integrated bin inventory, $w_i=1$. Quantities defined on different numerical
supports are conservatively projected onto common physical bins before
comparison.
For integrated inventories, $E_1$ quantifies relative disagreement in total
amount, while the signed relative difference indicates over- or
underprediction. Streamwise $E_1$ quantifies disagreement in the axial
inventory distribution while retaining disagreement in total amount. Spatial
$E_1$ quantifies disagreement in the two-dimensional inventory distribution at
the common-bin scale.

For the two-scale streamwise profiles, we additionally report the
independently normalized shape discrepancy,
\[
    L_{1,\mathrm{shape}} =
    100\sum_b\left|
    \frac{M_b^{\mathrm{DBS}}}{\sum_b M_b^{\mathrm{DBS}}}
    -
    \frac{M_b^{\mathrm{ref}}}{\sum_b M_b^{\mathrm{ref}}}
    \right|,
\]
where $M_b$ is the inventory in a common streamwise bin. This measure
compares the axial distribution after removing differences in total
inventory; the signed inventory differences are reported separately.

\subsection{1D Retardation Benchmark}\label{sec:1D}

The 1D linear-retardation benchmark tests the surface-chemistry coupling on a
uniform domain where the retardation factor has an exact analytical form.
With uniform mobile-water fraction $\theta$ and surface-area density
$a_{\mathrm{surf}}$, the retardation coefficient is
$R = 1 + a_{\mathrm{surf}} K_s / \theta$, and the retarded concentration front
satisfies the advection--diffusion equation with effective velocity
$U_D/(\theta R)$ and effective diffusivity $D/R$. This configuration isolates
the batch calibration of $K_s$, the cell-based surface-capacity term, and the
advective--diffusive transport step. It compares a surface-enabled transport
case, a matched aqueous-only case ($a_{\mathrm{surf}}=0$), and the dilute batch
pair used to estimate $K_s$ by a secant between two equilibrium states.

The retained batch pair gives $K_s=4.253\times10^{-6}$~m and $R=1.851$, so the
surface-enabled front travels at $1/R\approx54\%$ of the aqueous-only front
velocity. The transport mesh contains $80\times1\times1$ cells and uses a
species diffusion coefficient of $10^{-11}\ \mathrm{m^2\,s^{-1}}$. Spatial
profiles at $t=10$~s are compared with the classical semi-infinite
Ogata--Banks solution~\citep{ogata1961solution}. Outlet breakthrough is
compared with a high-resolution finite-domain reference evaluated with the
same fixed-concentration inlet and zero-gradient outlet as the OpenFOAM case.

The mesh- and timestep-refinement study in Online Resource~1 shows
approximately first-order convergence. Relative to the finest combined result,
the reported discretisation gives profile differences of $E_1=3.01\%$ and
$1.05\%$ for the surface-enabled and aqueous-only cases, respectively. These
differences quantify the combined spatial, temporal, and sequential-coupling
errors without conflating them with an unmatched outlet condition.

\begin{figure}[htbp]
    \centering
    \includegraphics[width=0.9\linewidth]{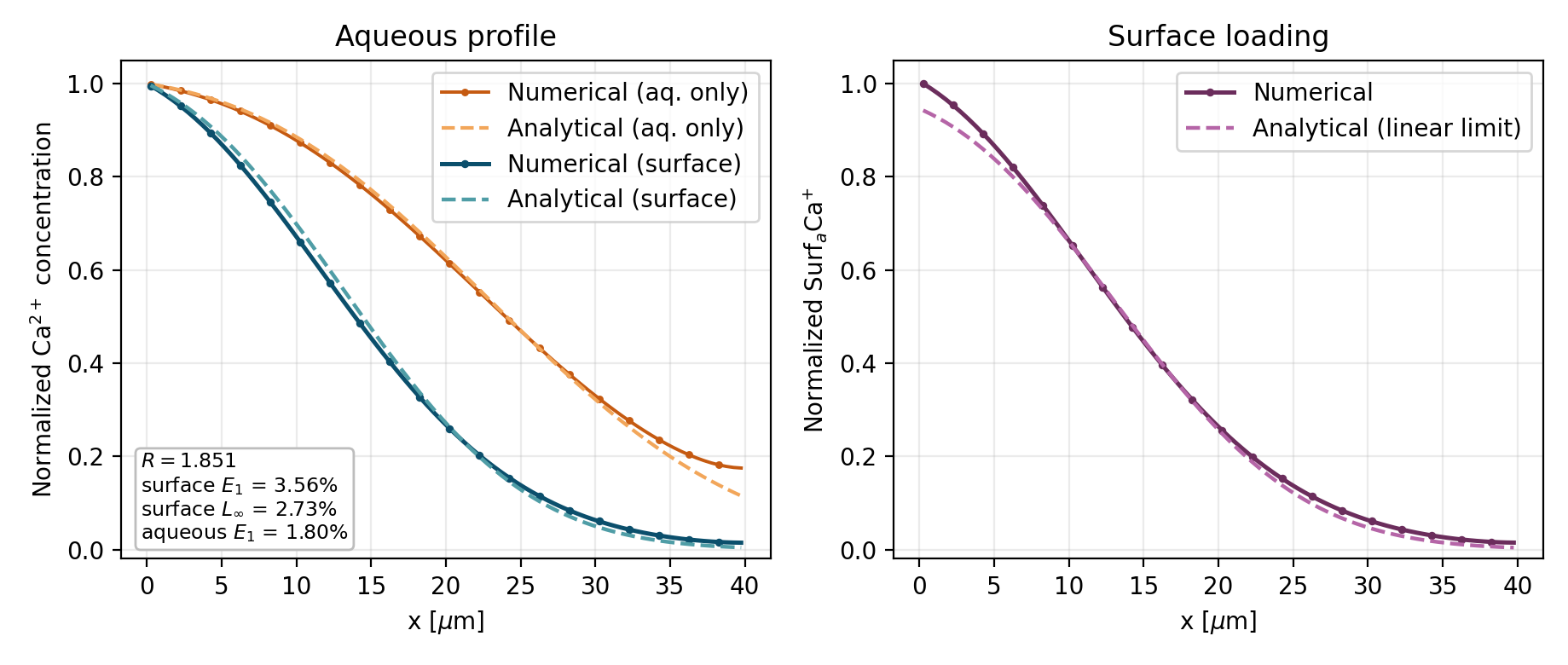}
    \caption{1D retardation benchmark: spatial profiles at $t=10$~s. Left:
    normalised aqueous Ca$^{2+}$ profiles for the aqueous-only and
    surface-enabled runs, compared with the corresponding cell-averaged
    semi-infinite Ogata--Banks solutions. Inset errors are measured relative
    to these semi-infinite profiles. Right: normalised adsorbed amount
    compared with the linear-equilibrium reference constructed from the
    batch-calibrated $K_s$ and the retarded Ogata--Banks profile.}
    \label{fig:1DRetardation}
\end{figure}

\begin{figure}[htbp]
    \centering
    \includegraphics[width=0.9\linewidth]{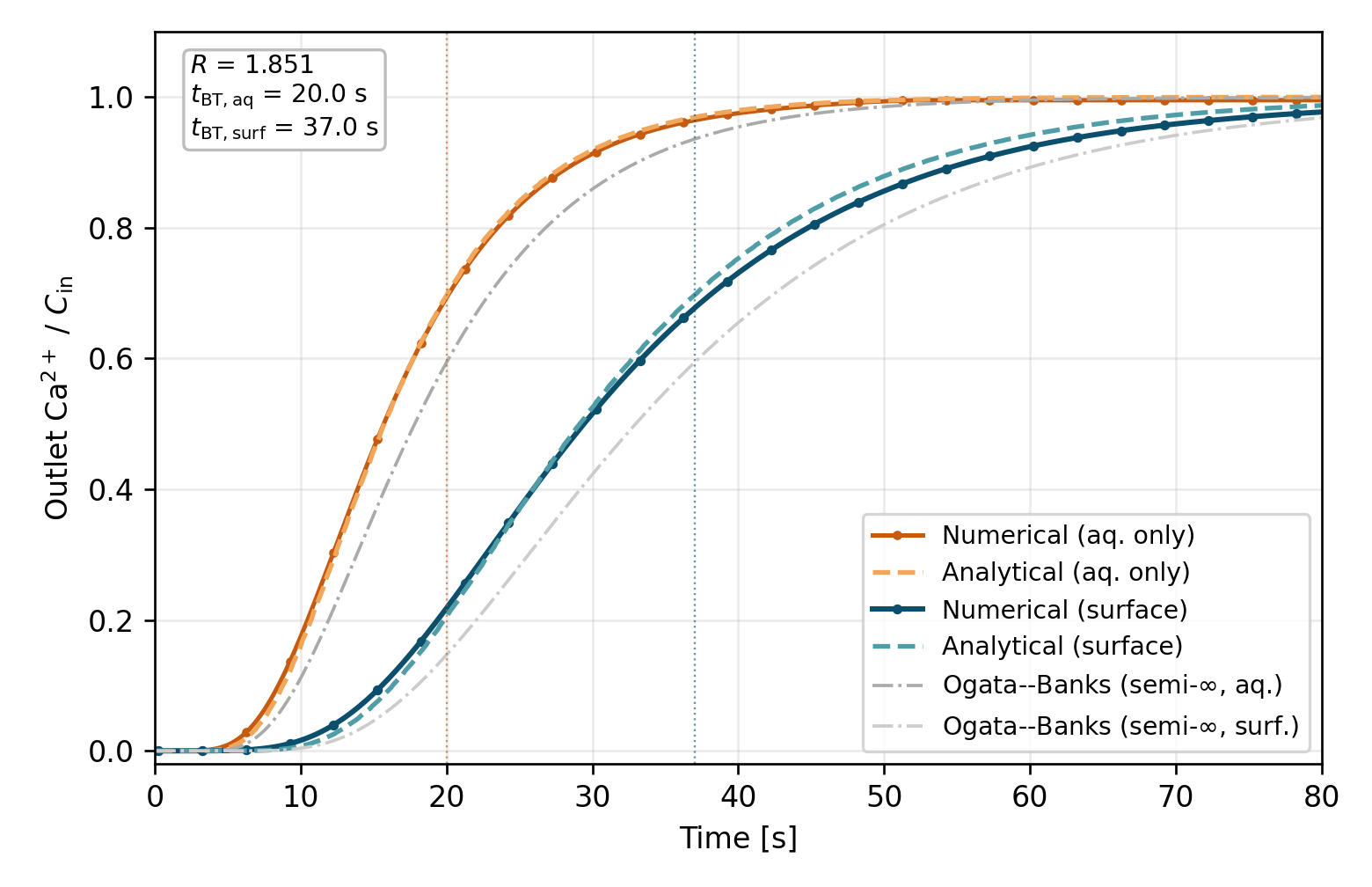}
    \caption{1D outlet breakthrough curves. Symbols show the outlet-patch
    area-averaged Ca$^{2+}$ normalised by the inlet concentration
    $C_{\mathrm{in}}$ for the aqueous-only and surface-enabled runs. Dashed
    lines are high-resolution finite-domain references with the same inlet and
    outlet conditions as the numerical cases. Dash-dot lines show the
    semi-infinite Ogata--Banks solutions; their difference from the dashed
    curves illustrates the effect of the finite-domain outlet condition.
    Vertical dotted lines mark the
    advective breakthrough times $t_{\mathrm{BT}}=L\theta/U_D$ (aqueous-only)
    and $t_{\mathrm{BT}}=L\theta R/U_D$ (surface-enabled).}
    \label{fig:1DBreakthrough}
\end{figure}

Figure~\ref{fig:1DRetardation} shows the spatial profiles at $t=10$~s, and
Figure~\ref{fig:1DBreakthrough} shows outlet breakthrough to $t=80$~s. The
surface-enabled front travels at $1/R$ of the aqueous-only velocity, giving
advective breakthrough times of 37.0 and 20.0~s, respectively, consistent with
the calibrated retardation factor. The finite-domain reference captures the
earlier outlet rise caused by the zero-gradient boundary. The semi-infinite
Ogata--Banks curves are also shown: their difference from the finite-domain
reference isolates the effect of the outlet boundary condition.

\FloatBarrier

\subsection{Closed-System Calibration and Batch Balance}

The closed-batch pair serves two purposes. Its two equilibrium states provide
the secant coefficient $K_s=\Delta\Gamma/\Delta C$ used throughout the 1D
benchmark, and each state tests whether one OpenFOAM--PhreeqcRM chemistry update
preserves Ca while repartitioning it between the aqueous and cell-based surface
stores. With no flow, diffusion, or boundary flux, conservation requires
\begin{equation}
    \mathcal{M} = \theta C + a_{\mathrm{surf}}\Gamma = \mathrm{const},
    \label{eq:closedBalance}
\end{equation}
where $C$ is the aqueous concentration and $\Gamma$ is the adsorbed amount per
reactive area.
Let $C_0$ denote the aqueous concentration at the start of the chemistry update
and $\Gamma_0$ the corresponding initial adsorbed amount per reactive area. In a
consistent SI basis, $C$ is in mol\,m$^{-3}$, $\Gamma$ in mol\,m$^{-2}$, and
$a_{\mathrm{surf}}\Gamma$ in mol\,m$^{-3}$ of bulk volume. OpenFOAM stores the
same fields in kmol\,m$^{-3}$ and kmol\,m$^{-2}$; multiplying both amount fields
by $10^3$ leaves the balance unchanged, and the numerical value of kmol\,m$^{-3}$
is identical to mol\,L$^{-1}$. In the dilute linear limit,
Eq.~\eqref{eq:closedBalance} gives
\begin{equation}
    C_{\mathrm{eq}} =
    \frac{\theta C_0+a_{\mathrm{surf}}\Gamma_0}
         {\theta+a_{\mathrm{surf}}K_s}.
    \label{eq:batchEquil}
\end{equation}

The two calculations use an $80\times1\times1$ mesh, $\theta=0.5$,
$a_{\mathrm{surf}}=10^5\ \mathrm{m^{-1}}$, and the same Ca$^{2+}$ chemistry on
\texttt{Surf\_a} as the transport benchmark. The scripted initial aqueous
values are $10^{-9}$ and $2\times10^{-9}$~kmol\,m$^{-3}$, numerically equal to
mol\,L$^{-1}$ and equivalent to $10^{-6}$ and $2\times10^{-6}$~mol\,m$^{-3}$
on the SI basis of Eq.~\eqref{eq:closedBalance}. Before the chemistry update,
the surface is initialised from the dilute \texttt{SOLUTION 1} state, giving
$\Gamma_0=1.132\times10^{-19}$~mol\,m$^{-2}$. The retained equilibrium pair
gives $K_s=4.253\times10^{-6}$~m.
Using $\mathcal{M}_0=\theta C_0+a_{\mathrm{surf}}\Gamma_0$, the signed relative
inventory changes are $+8.9\times10^{-10}$ and $-1.1\times10^{-10}$; the maximum
magnitude is below $10^{-9}$. These conservation residuals are calculated from
fields written to 16-digit precision. The test detects errors in the
water-volume, surface-area, amount-unit, and field-mapping conversions across
the OpenFOAM--PhreeqcRM interface, any of which can create or remove field-level
Ca inventory even when PHREEQC executes successfully.

\subsection{Controlled Surface-Area Validation}

The first geometric validation step isolates the construction of
$a_{\mathrm{surf}}$ from transport and chemistry by using analytic curved
interfaces. Two cases are used so that both dimensional branches of
\texttt{smoothSurface} are tested. The 3D branch is evaluated on a voxelised
sphere with reference area $4\pi R^2$; the quasi-2D branch is evaluated on a
circular post with reference area $2\pi R h$, where $h$ is the out-of-plane
thickness. These benchmarks test how closely each image-based surface-area
construction reproduces the intended smooth interface.

The smooth-interface reconstruction recovers a sub-voxel interface before
mapping its area back into the DBS cells. The figure reports four resolutions,
$R/\Delta x = 8$, 16, 32, and 64. For the 3D sphere, using a Gaussian occupancy
width of $\sigma=0.75$ voxels, the \texttt{smoothSurface} errors are
$-2.40\%$, $-0.06\%$, $+0.70\%$, and $+0.87\%$, respectively. The body-fitted
\texttt{snappyHexMesh} comparator remains within approximately $1\%$; its
errors at $R/\Delta x=16$ and 64 are $-0.65\%$ and $-0.50\%$.
For the quasi-2D circular post, using $\sigma=1.5$ voxels, the
\texttt{smoothSurface} contour-length error remains within $0.2\%$ across the
same four resolutions, while the largest \texttt{snappyHexMesh} error is below
$0.1\%$. The circular-cylinder and sphere tests involve different surface
curvatures and body-fitted discretisations, so their \texttt{snappyHexMesh}
errors need not coincide. Neither smooth-interface curve decreases
monotonically to zero.
These controlled tests therefore establish branch- and parameter-specific
bounded error over the plotted range, not convergence or a common accuracy
level for the quasi-2D and 3D constructions.

\begin{figure}[htbp]
    \centering
    \includegraphics[width=\linewidth]{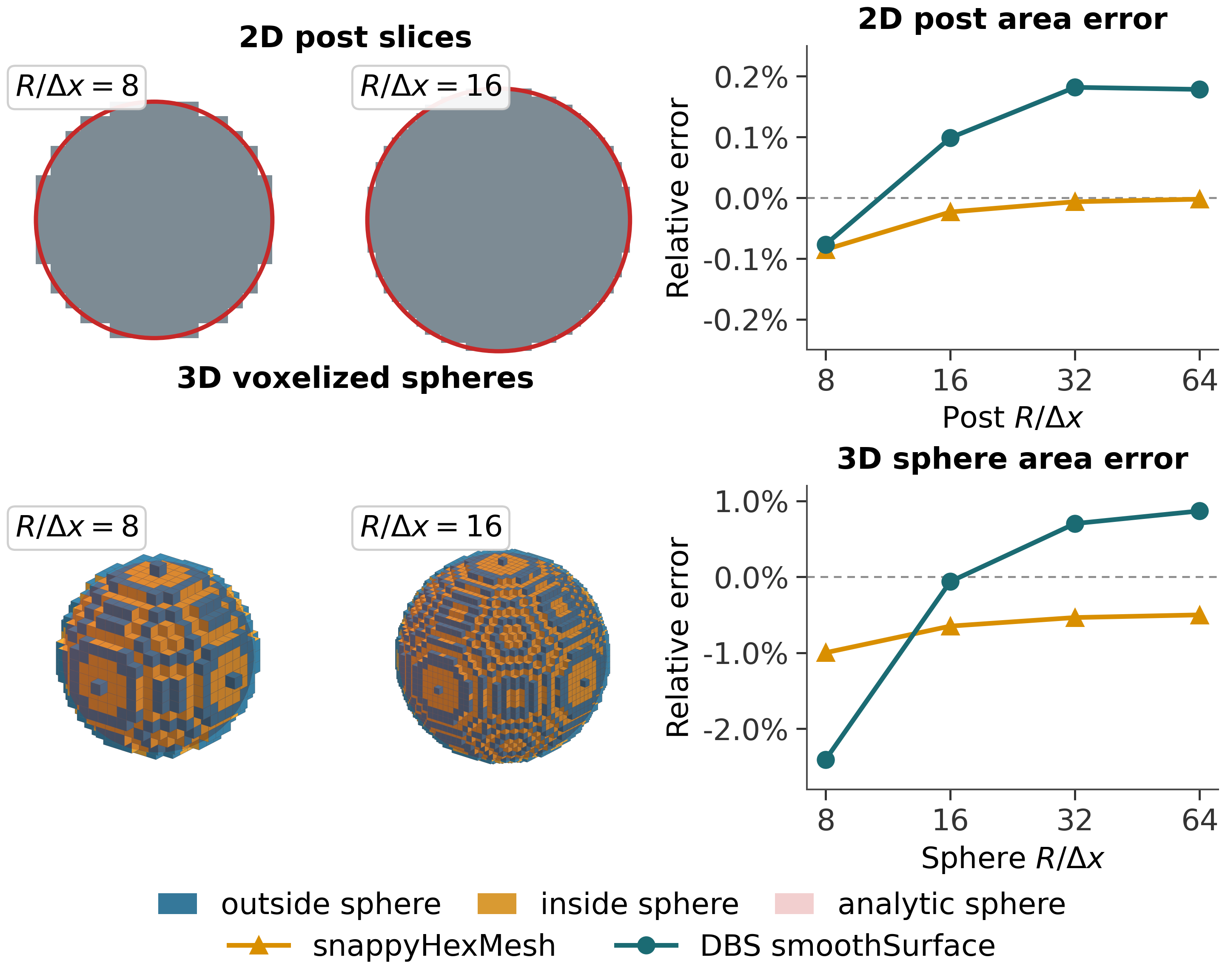}
    \caption{Controlled surface-area tests for the quasi-2D and 3D
    smooth-interface reconstructions. Left: voxelised circular-post slices,
    with the analytic circular interface shown in red, and three-dimensional
    renderings of the voxelised spheres at $R/\Delta x = 8$ and 16, overlaid
    by the semi-transparent red analytic spheres. Blue and amber identify voxel
    faces outside and inside the analytic sphere; these geometry colours are
    distinct from the method colours used in the error plots. Right: relative
    integrated surface-area error at $R/\Delta x=8$, 16, 32, and 64 for the 2D
    circular post ($A_{\mathrm{ref}} = 2 \pi R h$) and 3D sphere
    ($A_{\mathrm{ref}} = 4 \pi R^2$). The analytic areas define error;
    \texttt{snappyHexMesh} is a body-fitted comparator. The quasi-2D and 3D
    \texttt{smoothSurface} curves use $\sigma=1.5$ and $0.75$ voxels,
    respectively, and show bounded errors whose magnitudes do not decrease
    monotonically with refinement.}
    \label{fig:sphereSurfaceValidation}
\end{figure}

\FloatBarrier

\subsection{Micromodel Image-Based Surface-Area Comparison}

The image-based test moves from analytic geometry to a reactive micromodel for
which no exact surface-area reference exists. It compares two representations
of the same segmented geometry: a resolved-wall \texttt{snappyHexMesh}
calculation and DBS \texttt{smoothSurface}. The resolved-wall calculation is a
numerical comparator. The comparison tests whether the two surface
representations give similar integrated area and reactive-transport observables
after matching the inlet volumetric flux.

The resolved-wall and DBS results represent the image-derived geometry with
different numerical objects. In the \texttt{snappyHexMesh} case, the geometry
is a body-fitted wall and its face area is mapped to the adjacent chemistry
cells as an explicit surface-area density. In DBS, the same area is obtained
from the image-based smooth-interface reconstruction. Surface complexation in
both cases therefore samples a cell-supported area density together with the
local aqueous concentration. The pressure drops are calibrated separately to
match the inlet volumetric flux. The comparison is made on the integrated
reactive area, its spatial distribution, the aqueous transport profile, and the
resulting sorbed inventory. Because wall faces and Cartesian reconstruction
cells carry area on different supports, a sub-cell displacement can produce
order-unity local relative differences or a zero denominator even when nearby
interfaces and integrated areas agree. Local maps are therefore interpreted as
area redistribution, while totals and common streamwise bins provide the
quantitative comparison.

For the resolved-wall \texttt{snappyHexMesh} reference, the integrated
reactive area is $1.65 \times 10^{-5}$~m$^2$, compared with
$1.67 \times 10^{-5}$~m$^2$ for DBS \texttt{smoothSurface}, an error of
$+1.48\%$. The separately calibrated pressure drops give inlet volumetric
fluxes that agree within $0.004\%$. At $t=200$~s, the dissolved inventory
differs by $-0.07\%$ and the streamwise aqueous-inventory-profile discrepancy is
$E_1=5.11\%$.
The DBS sorbed inventory differs by $+1.68\%$. The integrated area, dissolved
inventory, streamwise profile, and sorbed inventory therefore quantify the
agreement between the two numerical representations; the local area
distributions remain support-dependent.

\begin{figure}[htbp]
    \centering
    \includegraphics[width=0.9\linewidth]{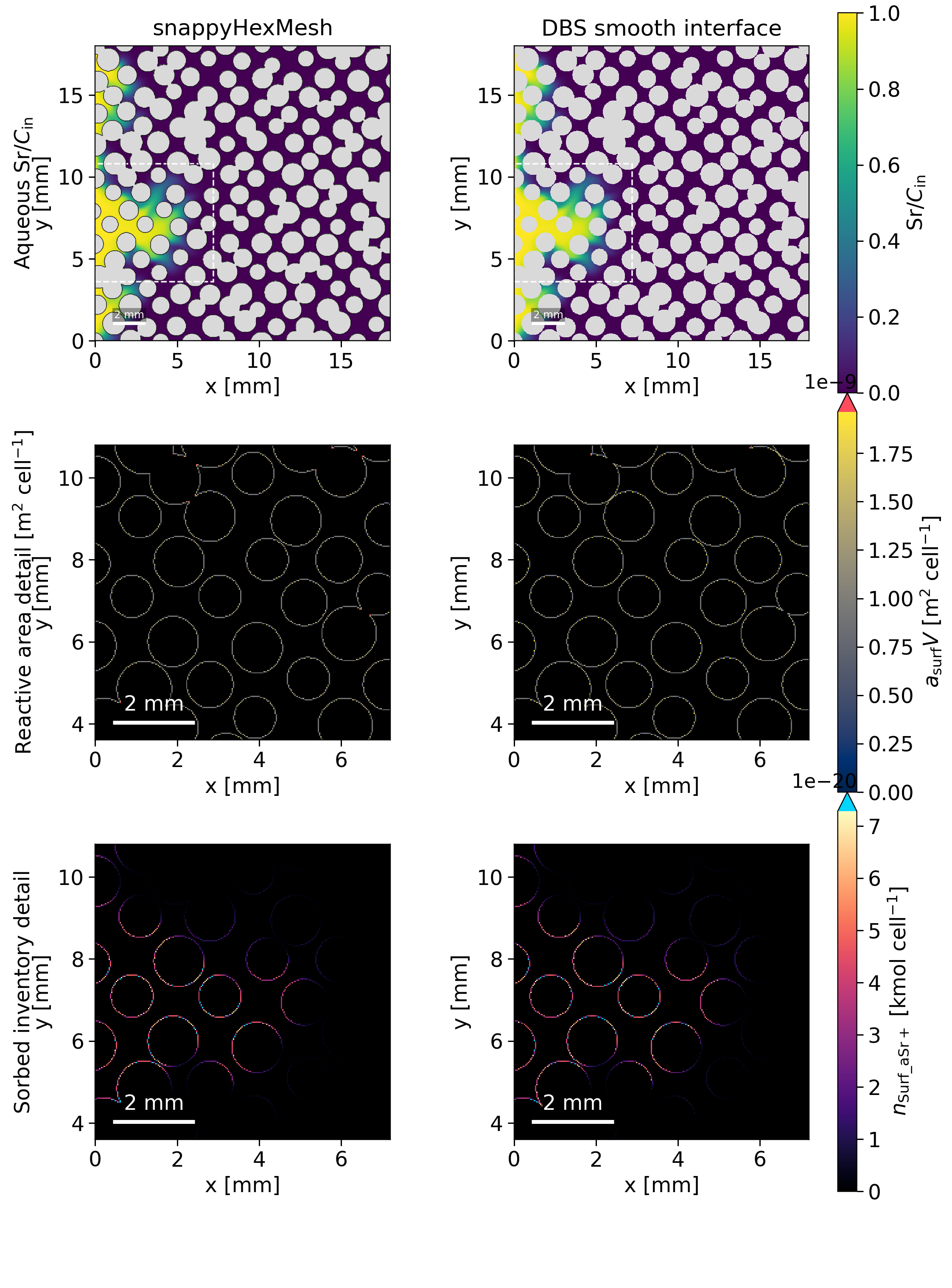}
    \caption{Micromodel image-based comparison at $t = 200$~s. The
    comparison is reported between the resolved-wall
    \texttt{snappyHexMesh} comparator and DBS \texttt{smoothSurface}. Rows show
    dissolved Sr$^{2+}$ fields normalised by
    $C_{\mathrm{in}}$, a magnified region of cell-wise reactive surface area
    $a_{\mathrm{surf}}V$, and the corresponding sorbed inventory fields over
    the same region; the dashed boxes in the top row mark the region shown at
    higher magnification. The lower rows show the chemistry-cell inventories
    for both cases; in the \texttt{snappyHexMesh} case, each wall face
    contributes its area to its adjacent cell. The area row reports both the
    integrated area and its spatial distribution, which controls the
    surface-complexation inventory.
    The two representations use the same nominal image resolution but place
    the reconstructed interface and its adjacent flow paths on different
    numerical supports.}
    \label{fig:hm12SurfaceValidation}
\end{figure}

Table~\ref{tab:micromodelProfileErrors} summarises the global errors for the
same comparison. The aqueous profile discrepancy is evaluated from aqueous
inventory conservatively accumulated into common streamwise bins.

\begin{table}[htbp]
    \centering
    \caption{Micromodel differences relative to the resolved-wall
    \texttt{snappyHexMesh} comparator at $t = 200$~s. Inventory and area
    differences are signed relative differences with \texttt{snappyHexMesh} as
    the numerical denominator; the aqueous profile discrepancy is the relative
    $L_1$ measure in Eq.~\eqref{eq:relativeL1}.}
    \label{tab:micromodelProfileErrors}
    \begin{tabular}{lcccc}
        \toprule
        Method & \shortstack{Dissolved\\inventory [\%]} & \shortstack{Surface\\area [\%]} & \shortstack{Sorbed\\inventory [\%]} & \shortstack{Aqueous\\profile\\$E_1$ [\%]} \\
        \midrule
        DBS \texttt{smoothSurface} & $-0.07$ & $+1.48$ & $+1.68$ & $5.11$ \\
        \bottomrule
    \end{tabular}
\end{table}

\FloatBarrier

\subsection{3D External Surface-Area and Transport Comparison}

The same resolved-wall/DBS method comparison is then applied to a
three-dimensional image crop. A $128^3$-voxel Estaillades subvolume is used,
with the micropore label remapped to solid so that the comparison isolates the
external surface between pore and grain. The two cases use the same
Sr$^{2+}$/Ca$^{2+}$ surface-complexation chemistry and separately calibrated
pressure drops to match the Darcy velocity: a resolved-wall
\texttt{snappyHexMesh} comparator and DBS \texttt{smoothSurface}. The
reactive-transport comparison is made at
$t = 2.5$~s, for which the advective front has traversed approximately half
of the crop under the resolved-wall Darcy velocity. For this external-surface
validation, the \texttt{smoothSurface} Gaussian occupancy width is
$\sigma = 0.74$ voxels.

The 3D comparison separates three quantities that affect the reactive result:
the DBS flow field, the integrated external area, and the sorbed inventory that
follows from their local overlap. The calibrated Darcy velocities agree within
$1.12\%$. Relative to the resolved-wall comparator, the
\texttt{smoothSurface} integrated reactive area is $1.52\%$ greater and the
sorbed Sr$^{2+}$ inventory is $1.84\%$ lower. These global differences quantify
agreement between the two numerical representations on their native supports.

\begin{figure}[htbp]
    \centering
    \includegraphics[width=0.895\linewidth]{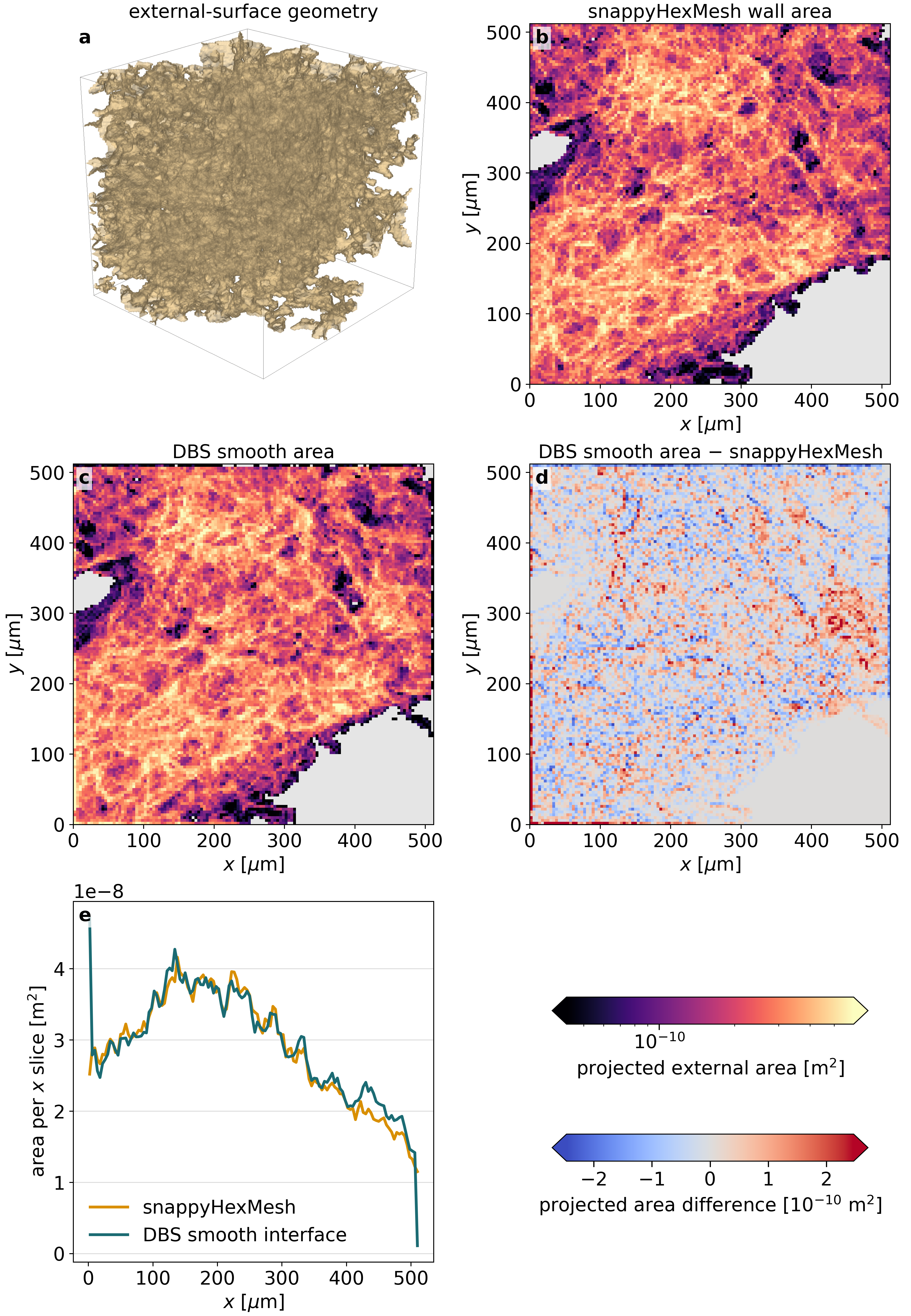}
    \caption{Three-dimensional external-surface comparison on the Estaillades
    crop. The 3D rendering shows the connected external surface between pore
    and grain after remapping the micropore label to solid. Projected
    surface-area maps compare the resolved-wall \texttt{snappyHexMesh}
    comparator and DBS \texttt{smoothSurface}. The signed projected-area
    difference map reports DBS minus \texttt{snappyHexMesh}, and the streamwise
    profiles compare area accumulated into common axial bins. The
    \texttt{snappyHexMesh} wall faces are binned onto the image grid only for
    these spatial comparisons; integrated areas and inventories are computed on
    their native numerical supports.}
    \label{fig:external3DSurfaceValidation}
\end{figure}

\begin{table}[htbp]
    \centering
    \caption{Three-dimensional external-surface comparison on the Estaillades
    crop at $t = 2.5$~s. Surface-area and sorbed Sr$^{2+}$ inventory
    differences are signed relative differences with the resolved-wall
    \texttt{snappyHexMesh} comparator as the numerical denominator. The
    \texttt{smoothSurface} case uses $\sigma = 0.74$ voxels.}
    \label{tab:external3DValidation}
    \begin{tabular}{lcc}
        \toprule
        Method & \shortstack{Surface\\area [\%]} & \shortstack{Sorbed Sr\\inventory [\%]} \\
        \midrule
        DBS \texttt{smoothSurface} & $+1.52$ & $-1.84$ \\
        \bottomrule
    \end{tabular}
\end{table}

\FloatBarrier

\subsection{Resolved-to-DBS Two-Scale Sorption Benchmark}

The unresolved-matrix component of the DBS model is tested by comparing an
upscaled DBS representation with an aligned, fully resolved micromodel based on
the multiscale images used in the dispersivity study of
\citet{maes2024dispersivity}. The resolved reference uses a high-resolution
crop of the micromodel in which the matrix pore space is explicitly
represented. The DBS calculation uses the corresponding lower-resolution
image of the same physical region, in which the large vugs remain resolved
while the matrix phase is assigned porosity, permeability, effective
dispersion, and intrinsic reactive surface area from the high-resolution
matrix representative elementary volume. The matrix permeability provides
the unresolved flow resistance; the effective permeability of the complete
domain also depends on the geometry and connectivity of the resolved vugs.
For the $6\times6$~mm micromodel examined by
\citet{maes2024dispersivity}, a matrix permeability of
$4.04\times10^{-12}$~$\mathrm{m^2}$ yielded a whole-domain permeability of
$5.10\times10^{-12}$~$\mathrm{m^2}$ in both the fully resolved and DBS
calculations, equal at the reported precision.

\begin{figure}[htbp]
    \centering
    \includegraphics[width=\linewidth]{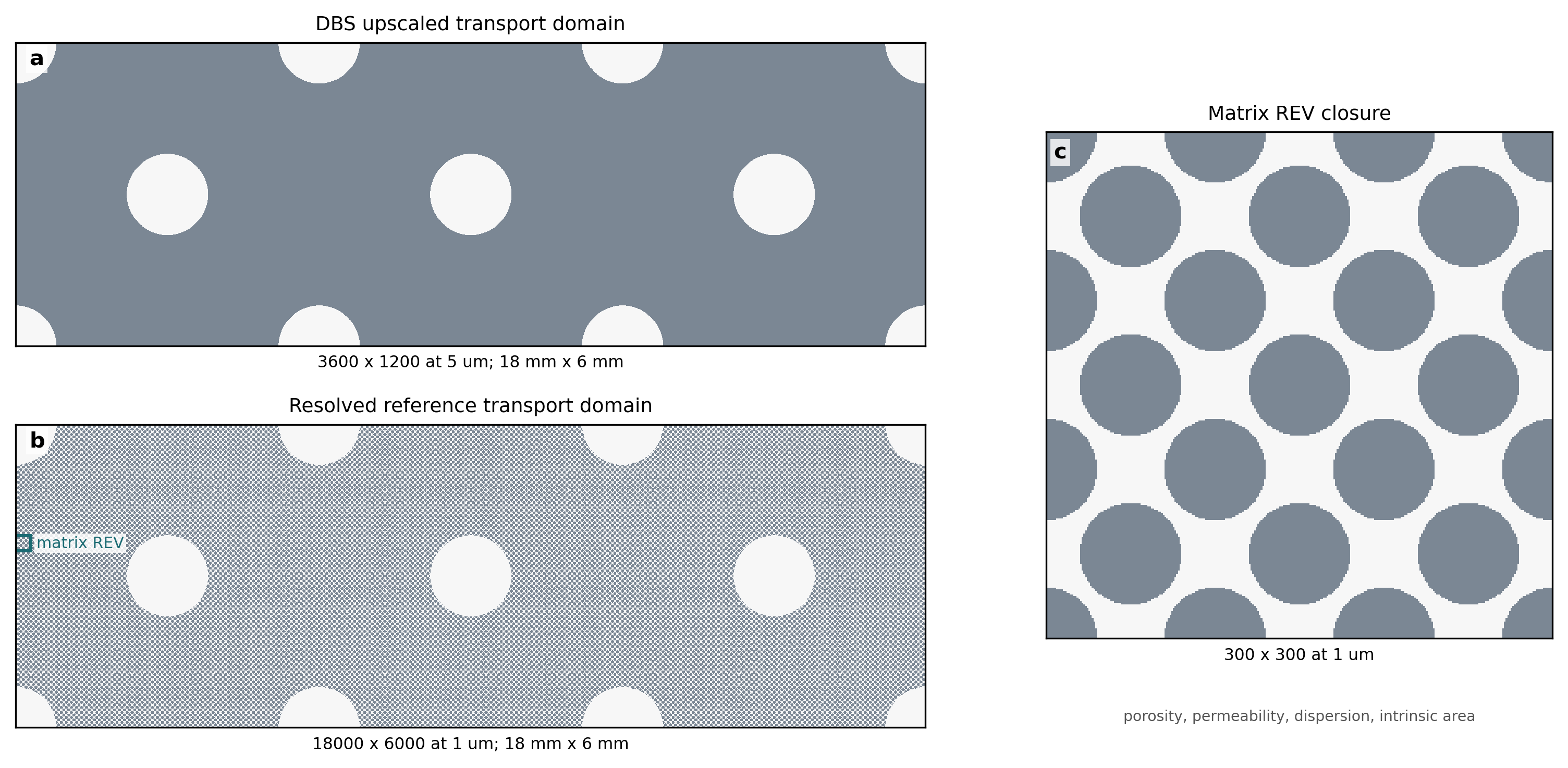}
    \caption{Nested geometry construction for the resolved-to-DBS two-scale
    micromodel benchmark. (a)~DBS upscaled transport domain, where the large
    vugs are resolved and the matrix phase is represented by averaged
    properties. (b)~Resolved reference transport domain over the same physical
    length and height, with the matrix pore space explicitly represented.
    (c)~High-resolution matrix representative elementary volume used to assign
    the unresolved matrix porosity, permeability, dispersivity, and intrinsic
    reactive surface area in the DBS calculation.}
    \label{fig:twoScaleLithiumNestedGeometries}
\end{figure}

Figure~\ref{fig:twoScaleLithiumNestedGeometries} shows the aligned geometry
hierarchy used for this comparison. The upscaled DBS and resolved-reference
transport domains have the same physical size, while the matrix representative
elementary volume provides the unresolved matrix properties used in the DBS
calculation.

\FloatBarrier

The reactive-transport problem is formulated as a critical-mineral recovery
analogue for direct lithium extraction. A Li-bearing Na-K-Mg brine invades an
initially Li-free brine, and Li$^{+}$ competes with the major brine cations for
a generic sorbent site:
\[
    \mathrm{Sor\_aH} + \mathrm{Li}^{+}
        \rightleftharpoons \mathrm{Sor\_aLi} + \mathrm{H}^{+}.
\]
Parallel reactions for Na$^{+}$, K$^{+}$, and Mg$^{2+}$ provide competing
surface occupancy. The chemistry is generic, representing a Li-selective
surface-complexation benchmark for engineered sorbent coatings rather than a
specific commercial direct-lithium-extraction material.

The resolved and DBS cases are simulated at matched pore-scale P\'{e}clet number,
\[
    Pe = \frac{U_D L_{\mathrm{pore}}}{D\theta},
\]
where $U_D$ is the Darcy velocity, $L_{\mathrm{pore}}$ is the pore length scale
reported by the flow post-processing, $D$ is the aqueous diffusion coefficient,
and $\theta$ is the porosity. Evaluating $Pe$ using the corresponding
$U_D$, $L_{\mathrm{pore}}$, $D$, and $\theta$ for each case places the
comparison at the same relative strength of advection and molecular diffusion
on the pore length scale. The field comparison shown below evaluates the
spatial Li$^{+}$ aqueous concentration and sorbed $\mathrm{Sor\_aLi}$ amounts
against the fully resolved reference at the same $Pe$ and matched advective
distance $x/L=U_Dt/(\theta L)=0.625$, where $t$ is the elapsed transport time
and $L$ is the common domain length. The quantitative $Pe$ sweep is evaluated
after 1.25 advective transits.

The P\'{e}clet number also controls the accuracy of the two-scale reduction. At
low $Pe$, diffusive exchange can equilibrate the
unresolved matrix over the residence time of the advecting fluid, so the DBS
cell behaves as an effective local reservoir with averaged porosity, diffusion
coefficient, permeability, and intrinsic surface area. At higher $Pe$, solute
can bypass part of the matrix before diffusion has accessed the internal
surface. In that regime, matching the integrated surface area is not sufficient:
the reactive prediction also depends on the time-dependent accessibility of
that area to the aqueous front. The resolved-to-DBS benchmark therefore tests
how well the two-scale reduction preserves surface accessibility as $Pe$
varies.

The choice of surface reconstruction was first tested in two short DBS
calculations at $Pe=1$. Over three matched checkpoints ending at
$0.99564208$~s, the reconstructions differ by at most $0.000647\%$ in
integrated area, $0.000703\%$ in aqueous Li inventory, and $0.2110\%$ in
sorbed Li inventory. The corresponding normalised L1 differences are
$0.00172\%$ and $0.5147\%$. This interval covers only $0.489\%$ of the
planned run, so the comparison establishes sensitivity to the area
reconstruction at early time.

\begin{figure}[htbp]
    \centering
    \includegraphics[width=\linewidth]{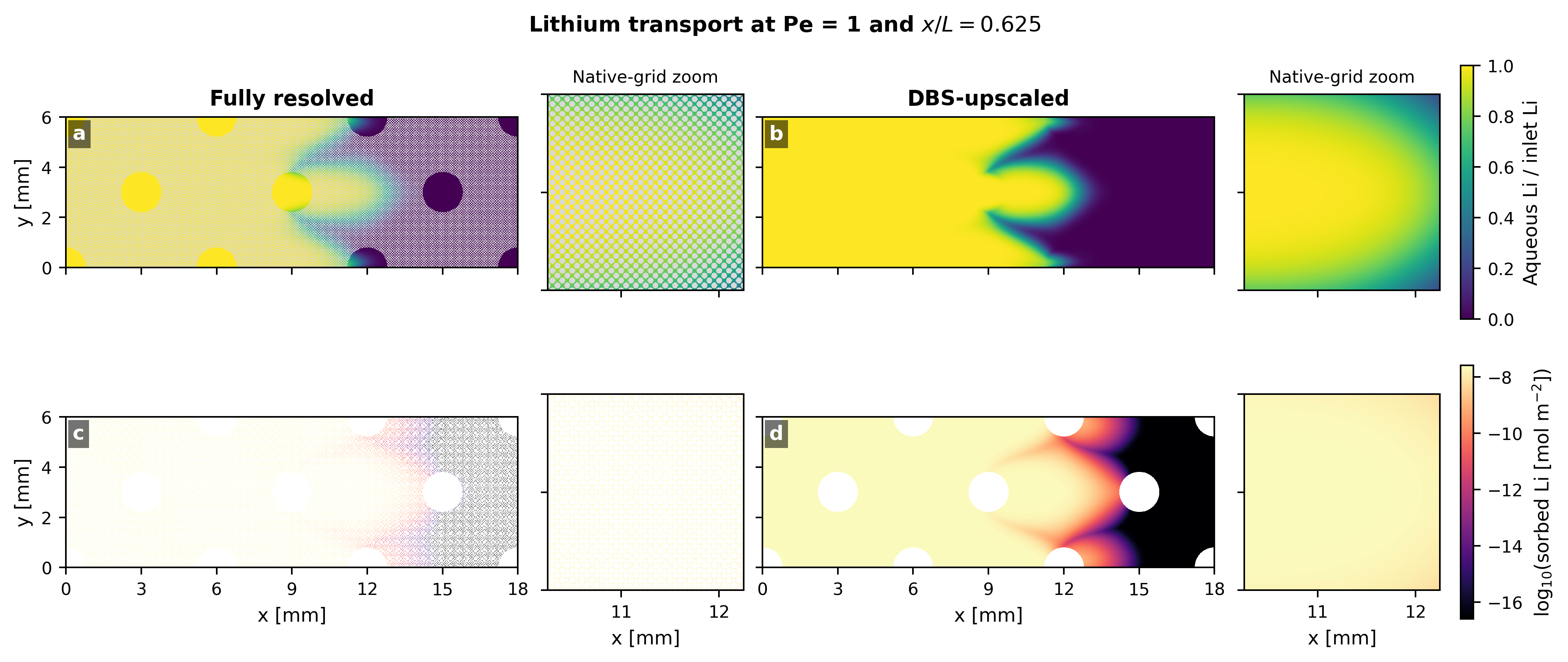}
    \caption{Native-grid lithium fields for the fully resolved and DBS-upscaled
    domains at $Pe=1$ and $x/L=0.625$. (a,b)~Aqueous Li concentration
    normalised by the inlet concentration. (c,d)~Sorbed Li inventory on a
    shared logarithmic scale. The adjacent zooms preserve each calculation's
    native $5~\mu$m grid; solid grains are masked in the resolved fields and no
    spatial averaging is applied for display.}
    \label{fig:twoScalePeOneFields}
\end{figure}

The native-grid fields at $Pe=1$ provide a direct spatial comparison
(Figure~\ref{fig:twoScalePeOneFields}). The DBS solution
reproduces the resolved calculation's advancing aqueous front and the
corresponding sorbed-Li distribution at $x/L=0.625$. The resolved geometry
retains pore-scale concentration variations around individual matrix grains,
whereas the DBS field represents their smooth cell-scale average.

The sweep covers $Pe=0.1$, 1, 3, and 10. Each resolved/DBS pair occupies the
same $18\times6$~mm domain on $5~\mu$m grids and is compared after 1.25
advective transits. Relative to the resolved model, DBS gives
$0.51$--$0.58\%$ less aqueous Li and $0.37$--$0.43\%$ more sorbed Li. The
streamwise aqueous profiles have
$L_{1,\mathrm{shape}}=0.54$--$0.58\%$, whereas the sorbed profiles have
$L_{1,\mathrm{shape}}=4.74$--$4.75\%$. Uptake is therefore
redistributed slightly along the flow path even as its total amount remains
close to the resolved result: the largest offsets in cumulative-inventory
median position are 5~$\mu$m in solution and 15~$\mu$m on the surface. For this geometry and
chemistry, the specified intrinsic-area closure reproduces domain-scale uptake
across the tested $Pe$ range.

\begin{figure}[htbp]
    \centering
    \includegraphics[width=\linewidth]{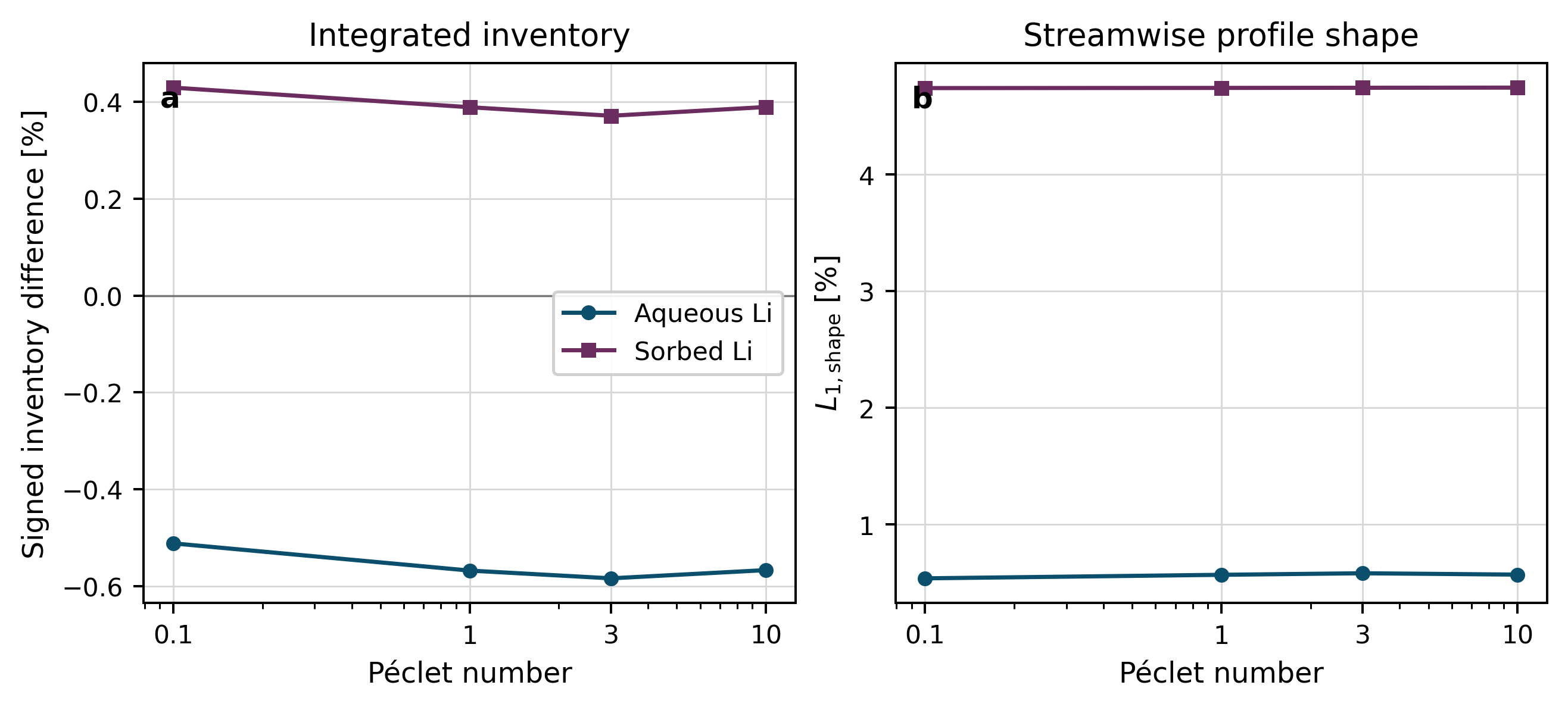}
    \caption{Resolved-to-DBS lithium-sorption comparison after 1.25 advective
    transits for $Pe=0.1$, 1, 3, and 10. Inventory differences are signed
    DBS-minus-resolved percentages. $L_{1,\mathrm{shape}}$ compares
    independently normalized inventories on common streamwise supports.}
    \label{fig:twoScaleErrorVsPe}
\end{figure}

\FloatBarrier

\subsection{Validation Summary}\label{sec:valSummary}

The validation sequence separates conservative aqueous--surface
state transfer, coupled transport, image-derived surface geometry, and
two-scale DBS sorption upscaling. The closed-batch pair tests the field-level Ca
balance during local equilibration, whereas the transient 1D benchmark tests
the retarded transport response and its mesh and timestep convergence on a
domain with no geometric ambiguity. The controlled sphere and circular-post
benchmarks isolate the geometric accuracy of the interface-area construction
itself. The micromodel solid-grain and 3D
external-surface comparisons transfer that geometric question to reactive image
benchmarks by comparing DBS \texttt{smoothSurface} with resolved-wall
\texttt{snappyHexMesh} references.
The resolved-to-DBS micromodel benchmark tests the two-scale reactive-transport
representation used when part of an image is treated as an unresolved reactive
matrix. Table~\ref{tab:validationSummary} consolidates the quantitative tests;
the geometric validation results are reported in
Figures~\ref{fig:sphereSurfaceValidation}, \ref{fig:hm12SurfaceValidation},
and~\ref{fig:external3DSurfaceValidation}, with global errors in
Tables~\ref{tab:micromodelProfileErrors} and~\ref{tab:external3DValidation};
the two-scale transport results are reported in
Figures~\ref{fig:twoScalePeOneFields} and~\ref{fig:twoScaleErrorVsPe}.

\begin{table}[htbp]
    \centering
    \caption{Validation measures for analytical, interface, and two-scale
    comparisons. Signed inventory differences are DBS minus the corresponding
    reference. Profile discrepancies use $E_1$ for the 1D and image-micromodel
    comparisons and $L_{1,\mathrm{shape}}$ for the resolved-to-DBS Li comparison. The resolved-to-DBS Li results
    are measured after 1.25 advective transits, whereas the reconstruction
    sensitivity covers only its three early checkpoints.}
    \label{tab:validationSummary}
    \footnotesize
    \setlength{\tabcolsep}{3pt}
    \begin{tabular}{@{}p{0.27\linewidth}p{0.43\linewidth}p{0.22\linewidth}@{}}
        \toprule
        Check & Metric & Result \\
        \midrule
        1D retardation & Profile $E_1$ relative to the finest result (surface / aqueous) & $3.01\%$ / $1.05\%$ \\
        Closed-batch pair & Maximum absolute relative inventory change & $8.9\times10^{-10}$ \\
        Circular post & area error & $<0.2\%$ \\
        Sphere, $R/\Delta x=8/16/32/64$ & area error & $-2.40/-0.06/+0.70/+0.87\%$ \\
        Image micromodel & area / sorbed / profile $E_1$ & $+1.48/+1.68/5.11\%$ \\
        3D external surface & area / sorbed inventory & $+1.52/-1.84\%$ \\
        Short DBS sensitivity & maximum sorbed inventory / L1 change & $0.2110/0.5147\%$ \\
        Resolved/DBS Li, $Pe=0.1$--10 & aqueous / sorbed inventory difference & $-0.51$ to $-0.58\%$ / $+0.37$ to $+0.43\%$ \\
        Resolved/DBS Li, $Pe=0.1$--10 & aqueous / sorbed $L_{1,\mathrm{shape}}$ & $0.54$--$0.58\%$ / $4.74$--$4.75\%$ \\
        \bottomrule
    \end{tabular}
\end{table}

The validation results therefore support the conservative coupling
and the smooth-interface construction of $a_{\mathrm{surf}}$ used in the
application cases that follow. Those cases then examine the additional
intrinsic micropore contribution in progressively richer image-based media.
\section{Test Cases}

Three image-based surface-complexation applications are considered here, each
exercising nonlinear multi-species coupling beyond a single retardation factor
and representing a different subsurface setting:
(i)~Pb$^{2+}$ sorption on a two-site Dzombak-Morel hydrous ferric oxide (HFO)
coating for the 2D four-fracture network case, representative of heavy-metal
retardation in fractured rock systems and permeable reactive barrier
remediation;
(ii)~selective Mn$^{2+}$ uptake from Mg$^{2+}$-rich brine on a two-site HFO
surface for the Kelvin-cell 3D case, representative of critical-metal capture
in microporous industrial sorbent media; and
(iii)~Sr$^{2+}$ leachate invading Ca$^{2+}$-bearing limestone formation water
for the Estaillades 3D case, representative of radionuclide migration from a
deep geological repository (DGR) for radioactive waste.
Each case applies the coupling, the image-derived interface area, and the
two-scale DBS reactive surface validated in Section~\ref{sec:validation}, and
in each the intrinsic micropore term in $a_{\mathrm{surf}}$
(Section~\ref{sec:aSurf}) adds to the uptake carried by the resolved interface.
Table~\ref{tab:simParams} lists the simulation parameters for the three
application cases together with the 1D retardation benchmark, and
Table~\ref{tab:chemParams} lists their geochemical parameters.
In all cases cells with $\theta \le \theta_{\mathrm{mask}} = 10^{-3}$ are
treated as chemically inactive, and PHREEQC thermodynamic data are drawn from
the \texttt{GeoChem.dat} database distributed with GeoChemFoam.
The 1D benchmark is presented in Section~\ref{sec:1D} and uses a prescribed
uniform Darcy velocity $U_D = 10^{-6}\ \mathrm{m\,s^{-1}}$; the 2D and 3D
image-based cases use a velocity field from a \texttt{simpleDBSFoam} flow
calculation with kinematic pressure drop
$\Delta p = 5\times10^{-5}\ \mathrm{m^{2}\,s^{-2}}$ and kinematic viscosity
$\nu = 10^{-6}\ \mathrm{m^{2}\,s^{-1}}$.

\begin{table}[htbp]
    \centering
    \caption{Simulation parameters for the three image-based application cases
    and the 1D retardation benchmark of Section~\ref{sec:1D}. Image dimensions
    are given in voxels and the voxel spacing $\delta x$ sets the physical scale
    of each model. Phase labels are integers in the segmented image; (r)
    follows the name of the reactive phase receiving the phase-volume end
    member $s_k$, whose cellwise intrinsic contribution is $f_k s_k$. The 1D
    case instead uses a prescribed uniform $a_{\mathrm{surf}}$ in a synthetic
    domain with no image source.
    The 2D four-fracture network case uses 8 Message Passing Interface (MPI)
    ranks
    ($4\times2\times1$ decomposition) and the Estaillades case uses 64 MPI ranks
    ($4\times4\times4$ decomposition); the 1D case is serial.}
    \label{tab:simParams}
    \scriptsize
    \setlength{\tabcolsep}{2pt}
    \begin{tabular}{@{}p{0.13\linewidth}p{0.09\linewidth}p{0.06\linewidth}p{0.09\linewidth}p{0.16\linewidth}p{0.08\linewidth}p{0.10\linewidth}p{0.07\linewidth}p{0.07\linewidth}@{}}
        \toprule
        Case
            & Image [vox]
            & $\delta x$ ($\mu$m)
            & Comp.\ grid
            & Phase labels
            & $D$ (m$^{2}$\,s$^{-1}$)
            & Reactive-area input (m$^{-1}$)
            & $\Delta t$ (s)
            & $t_{\mathrm{end}}$ (s) \\
        \midrule
        1D retardation
            & n/a
            & n/a
            & $80\times1\times1$
            & \mbox{bulk ($\theta=0.5$)}
            & $10^{-11}$
            & $10^{5}$
            & $0.25$
            & $80$ \\
        2D fractures (Pb$^{2+}$/HFO)
            & $500\times500\times1$
            & $100$
            & $250\times250\times1$
            & 1=\mbox{micropore (r)}, 2=pore
            & $10^{-9}$
            & $10^{5}$
            & $0.25$
            & $20000$ \\
        3D Kelvin (Mn$^{2+}$/Mg$^{2+}$)
            & $200^{3}$
            & $10$
            & $200^{3}$
            & 1=pore, 2=\mbox{micropore (r)}
            & $10^{-9}$
            & $10^{5}$
            & $4$
            & $120$ \\
        3D Estaillades (Sr$^{2+}$/Ca$^{2+}$)
            & $500^{3}$
            & $4$
            & $500^{3}$
            & 1=pore, 2=\mbox{micropore (r)}, 3=solid
            & $10^{-9}$
            & $10^{5}$
            & $2$
            & $120$ \\
        \bottomrule
    \end{tabular}
\end{table}

\begin{table}[htbp]
    \centering
    \caption{Geochemical and physical parameters for the three image-based
    application cases and the 1D retardation benchmark of
    Section~\ref{sec:1D}.
    All surface reactions are written in the form
    $\mathrm{SiteOH} + \mathrm{M}^{n+} \rightleftharpoons \mathrm{SiteOM}^{(n-1)+} + \mathrm{H}^+$,
    so positive $\log K$ indicates sorption that is thermodynamically
    favourable at neutral pH relative to proton release.
    The two-site HFO reaction set and $\log K$ values for Pb$^{2+}$,
    Mn$^{2+}$, and Mg$^{2+}$ follow \citet{dzombak1990surface}; the site
    densities used in the simulations are listed in the table.
    \texttt{Surf\_a} equilibrium constants for Ca$^{2+}$ and
    Sr$^{2+}$ are from the \texttt{GeoChem.dat} database distributed with
    GeoChemFoam \citep{maes2021geochemfoam,parkhurst2013phreeqc}.
    The diffusivity $D = 10^{-9}$\,m$^2$\,s$^{-1}$ is a representative
    free-aqueous value for divalent cations at 25\,\textdegree{}C; the reduced
    value $D = 10^{-11}$\,m$^2$\,s$^{-1}$ for the 1D benchmark is chosen to
    give a P\'{e}clet number of order unity over the 80-cell domain so that
    both advective and diffusive spreading are visible in the spatial profiles.
    In the 1D case, $10^{5}$\,m$^{-1}$ is the prescribed uniform cell field
    $a_{\mathrm{surf}}$. In each image case, $10^{5}$\,m$^{-1}$ is instead the
    microporous-phase end member $s_{\mathrm{micro}}$, and the local intrinsic
    field is $f_{\mathrm{micro}}s_{\mathrm{micro}}$; see
    Section~\ref{sec:aSurf}.}
    \label{tab:chemParams}
    \scriptsize
    \setlength{\tabcolsep}{2pt}
    \begin{tabular}{@{}p{0.13\linewidth}p{0.19\linewidth}p{0.19\linewidth}p{0.19\linewidth}p{0.19\linewidth}@{}}
        \toprule
        Parameter
            & 1D retardation
            & 2D Pb$^{2+}$/HFO
            & 3D Mn$^{2+}$/Mg$^{2+}$
            & 3D Sr$^{2+}$/Ca$^{2+}$ \\
        \midrule
        Surface model
            & \texttt{Surf\_a} (single-site)
            & Dzombak-Morel HFO (two-site)
            & Dzombak-Morel HFO (two-site)
            & \texttt{Surf\_a} (single-site) \\
        Reactive species
            & Ca$^{2+}$
            & Pb$^{2+}$
            & Mn$^{2+}$, Mg$^{2+}$
            & Sr$^{2+}$, Ca$^{2+}$ \\
        Site density $N_k$ (mol\,m$^{-2}$)
            & $2.4\times10^{-6}$
            & Hfo$_s$: $2\times10^{-4}$;\enspace Hfo$_w$: $6\times10^{-3}$
            & Hfo$_s$: $2\times10^{-4}$;\enspace Hfo$_w$: $6\times10^{-3}$
            & $2.4\times10^{-6}$ \\
        Surface $\log K$
            & Ca: $-5.85$
            & Pb/Hfo$_s$: $+4.65$;\enspace Pb/Hfo$_w$: $+0.30$
            & Mn/Hfo$_s$: $-0.40$;\enspace Mn/Hfo$_w$: $-3.50$;\enspace Mg/Hfo$_w$: $-4.60$
            & Ca: $-5.85$;\enspace Sr: $-5.5$ \\
        Diffusivity $D$ (m$^2$\,s$^{-1}$)
            & $10^{-11}$
            & $10^{-9}$
            & $10^{-9}$
            & $10^{-9}$ \\
        Uniform cell water fraction $\theta$
            & $0.5$
            & n/a
            & n/a
            & n/a \\
        Micropore end-member porosity $\phi_{\mathrm{micro}}$
            & n/a
            & $0.4$
            & $0.4$
            & $0.4$ \\
        Reactive-area input (m$^{-1}$)
            & $a_{\mathrm{surf}}=10^{5}$
            & $s_{\mathrm{micro}}=10^{5}$
            & $s_{\mathrm{micro}}=10^{5}$
            & $s_{\mathrm{micro}}=10^{5}$ \\
        $\theta_{\mathrm{mask}}$
            & $10^{-3}$
            & $10^{-3}$
            & $10^{-3}$
            & $10^{-3}$ \\
        $C_{\mathrm{in}}$ (mol\,L$^{-1}$)
            & Ca: $2\times10^{-9}$
            & Pb: $10^{-3}$
            & Mn: $10^{-5}$;\enspace Mg: $10^{-3}$
            & Ca: $0$;\enspace Sr: $10^{-4}$ \\
        $C_{\mathrm{init}}$ (mol\,L$^{-1}$)
            & Ca: $0$
            & $\approx 0$
            & Mn: $0$;\enspace Mg: $10^{-3}$
            & Ca: $10^{-3}$;\enspace Sr: $0$ \\
        pH (inlet / initial)
            & 7
            & 7
            & 7
            & 7 \\
        \bottomrule
    \end{tabular}
\end{table}

\FloatBarrier

\subsection{\texorpdfstring{2D Fracture Network: Pb$^{2+}$ on
Dzombak-Morel HFO}{2D Fracture Network: Pb2+ on Dzombak-Morel HFO}}

The 2D demonstration uses the GeoChemFoam benchmark fracture-network image
\nolinkurl{4StraightFractures}. The segmented domain has dimensions
$500 \times 500 \times 1$ with two labels corresponding to micropore ($1$) and
pore ($2$). The computational grid is $250 \times 250 \times 1$.
The four straight fractures give a simple flow field in which retardation by
iron-oxide coatings on the fracture walls can be followed cell by cell, as it
occurs in iron-rich fractured aquifers, acid mine drainage pathways, and
engineered permeable reactive barriers where HFO is the sorbent phase.

The chemistry represents equilibrium Pb$^{2+}$ sorption on HFO using the
Dzombak-Morel two-site model \citep{dzombak1990surface}. The strong-site
reaction has $\log K = +4.65$, compared with $+0.30$ for the weak-site
reaction, whereas weak sites are thirty times more abundant. Pb$^{2+}$ enters
a Pb-free domain at $C_{\mathrm{in}} = 10^{-3}$~mol~L$^{-1}$ and pH~7.

Flow is directed from the left inlet to the right outlet along $+x$, with
kinematic pressure fixed at $5\times10^{-5}$ and $0$~m$^2$\,s$^{-2}$ on the
respective faces. Velocity is zero-gradient at the inlet and outlet and
no-slip at the lateral walls. The inlet fixes Pb$^{2+}$, Cl$^{-}$, H$^{+}$,
and OH$^{-}$ to the inflow composition, while the dissolved fields are
zero-gradient at the outlet and lateral walls; the front and back patches are
empty. At pH~7 the HFO sites retard dissolved Pb$^{2+}$, which reaches most of
the domain by $t=20000$~s. The total reactive-area field combines the
reconstructed interface for micropore 1 $\rightarrow$ pore 2 with the
microporous-phase end member $s_{\mathrm{micro}}=10^{5}\ \mathrm{m^{-1}}$.
The label-1 phase has $\phi_{\mathrm{micro}}=0.4$, and mixed-cell contributions
are weighted by its geometric fraction $f_{\mathrm{micro}}$.

\begin{figure}[p]
    \centering
    \includegraphics[width=\linewidth]{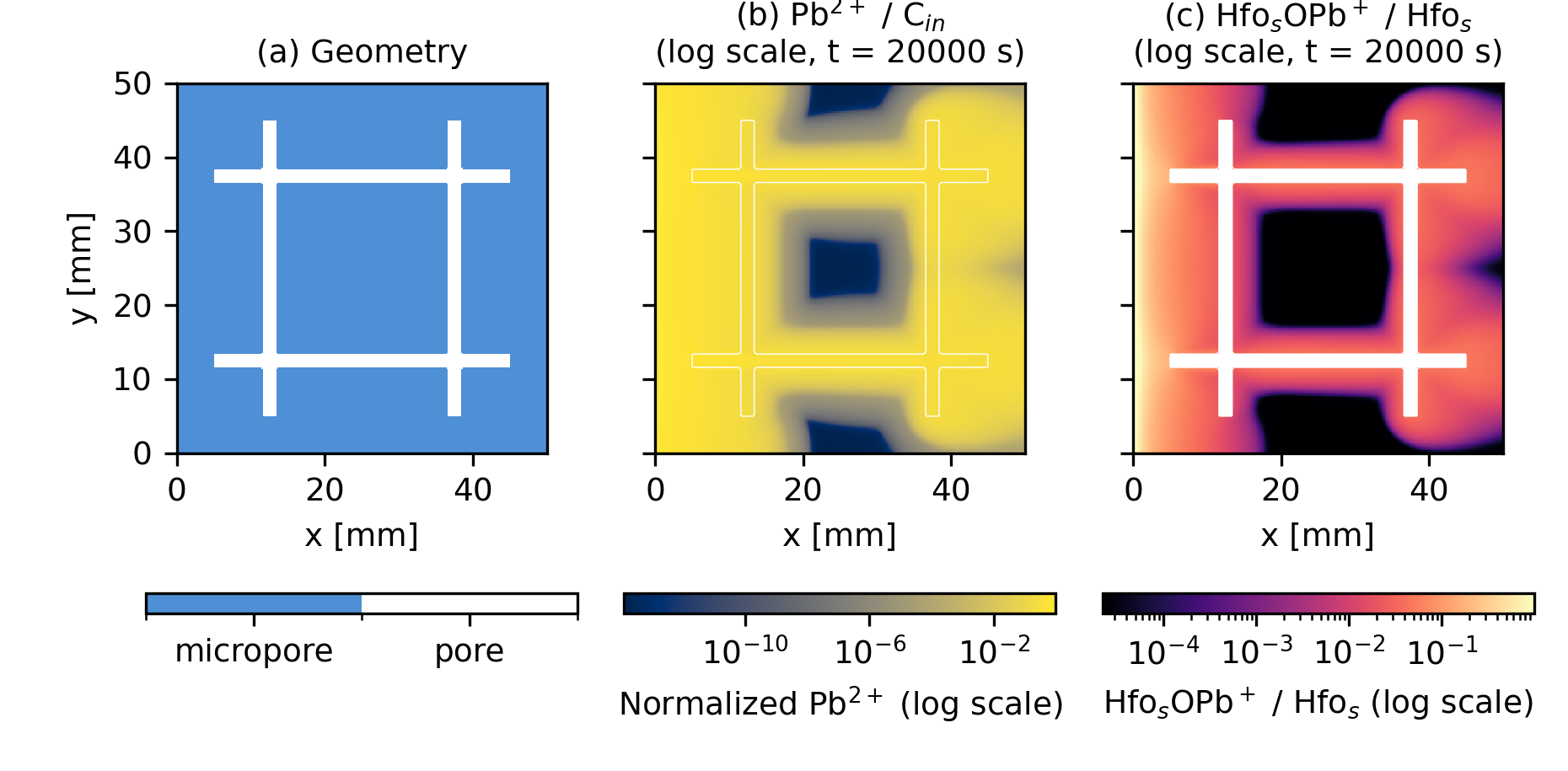}
    \caption{Dissolved Pb$^{2+}$ and strong-site occupancy in the 2D
    fracture-network case at $t=20000$~s. Left: segmented geometry, with
    microporous material in blue and open fracture space in white.
    Middle: Pb$^{2+}$/$C_{\mathrm{in}}$. Right:
    Hfo$_s$OPb$^+$/Hfo$_{s,\mathrm{tot}}$. Both are shown with logarithmic
    colour normalisation, and cells with zero reactive area are masked, leaving
    the open fracture geometry visible.}
    \label{fig:2DFractures}
\end{figure}

Figure~\ref{fig:2DFractures} shows the segmented geometry,
Pb$^{2+}/C_{\mathrm{in}}$, and the strong-site occupancy
Hfo$_s$\allowbreak OPb$^+$/\allowbreak Hfo$_{s,\mathrm{tot}}$ at $t=20000$~s. Dissolved Pb$^{2+}$
concentrations are higher along the open fracture channels than in the
surrounding microporous matrix. Strong-site occupancy is greatest near the
inlet and is locally enhanced in microporous material at fracture
intersections.

\FloatBarrier

\subsection{\texorpdfstring{3D Kelvin Cell: Selective Mn$^{2+}$
Capture from Mg$^{2+}$-Rich Brine}{3D Kelvin Cell: Selective Mn2+
Capture from Mg2+-Rich Brine}}

This case uses a Kelvin-cell image dataset with dimensions
$200 \times 200 \times 200$ and voxel size $10~\mu\mathrm{m}$, where
label $1$ denotes pore and label $2$ denotes a connected microporous lattice.
Transport is solved in DBS form through both labels; the micropore cells carry
$\theta = 0.4$ and the intrinsic reactive surface area.
Trace Mn$^{2+}$ is carried through a Mg$^{2+}$-rich background brine and the
two cations compete for the same HFO sites, as they do when Mn$^{2+}$ is
recovered from a high-hardness process or mine-water stream in an open-cell
sorbent medium. Mn$^{2+}$ complexes more strongly with HFO than Mg$^{2+}$ does,
so breakthrough and surface loading are set both by advection through the
Kelvin-cell geometry and by competitive exchange on the reactive coating.

The chemistry uses the same Dzombak-Morel two-site HFO model as the 2D Pb
case, with Mn$^{2+}$ and Mg$^{2+}$ competing on weak sites and Mn$^{2+}$ also
complexing with strong sites:
\begin{align}
    \mathrm{Hfo_sOH} + \mathrm{Mn^{2+}} &\rightleftharpoons \mathrm{Hfo_sOMn^{+}} + \mathrm{H^{+}}, \quad \log K = -0.40 \\
    \mathrm{Hfo_wOH} + \mathrm{Mn^{2+}} &\rightleftharpoons \mathrm{Hfo_wOMn^{+}} + \mathrm{H^{+}}, \quad \log K = -3.50 \\
    \mathrm{Hfo_wOH} + \mathrm{Mg^{2+}} &\rightleftharpoons \mathrm{Hfo_wOMg^{+}} + \mathrm{H^{+}}, \quad \log K = -4.60
\end{align}
Mg$^{2+}$ is present at $10^{-3}$~mol~L$^{-1}$ both in the initial water and at
the inlet, and Mn$^{2+}$ enters at $10^{-5}$~mol~L$^{-1}$ into a domain that
starts free of it. The surface sites are first equilibrated with the Mg-bearing
background water, so incoming Mn$^{2+}$ must displace part of the Mg inventory
in order to load the sites. Site densities are as in the 2D case, and
$a_{\mathrm{surf}}^{\mathrm{intrinsic}} = 10^{5}$~m$^{-1}$ in the micropore phase.

\begin{figure}[htbp]
    \centering
    \includegraphics[width=\linewidth]{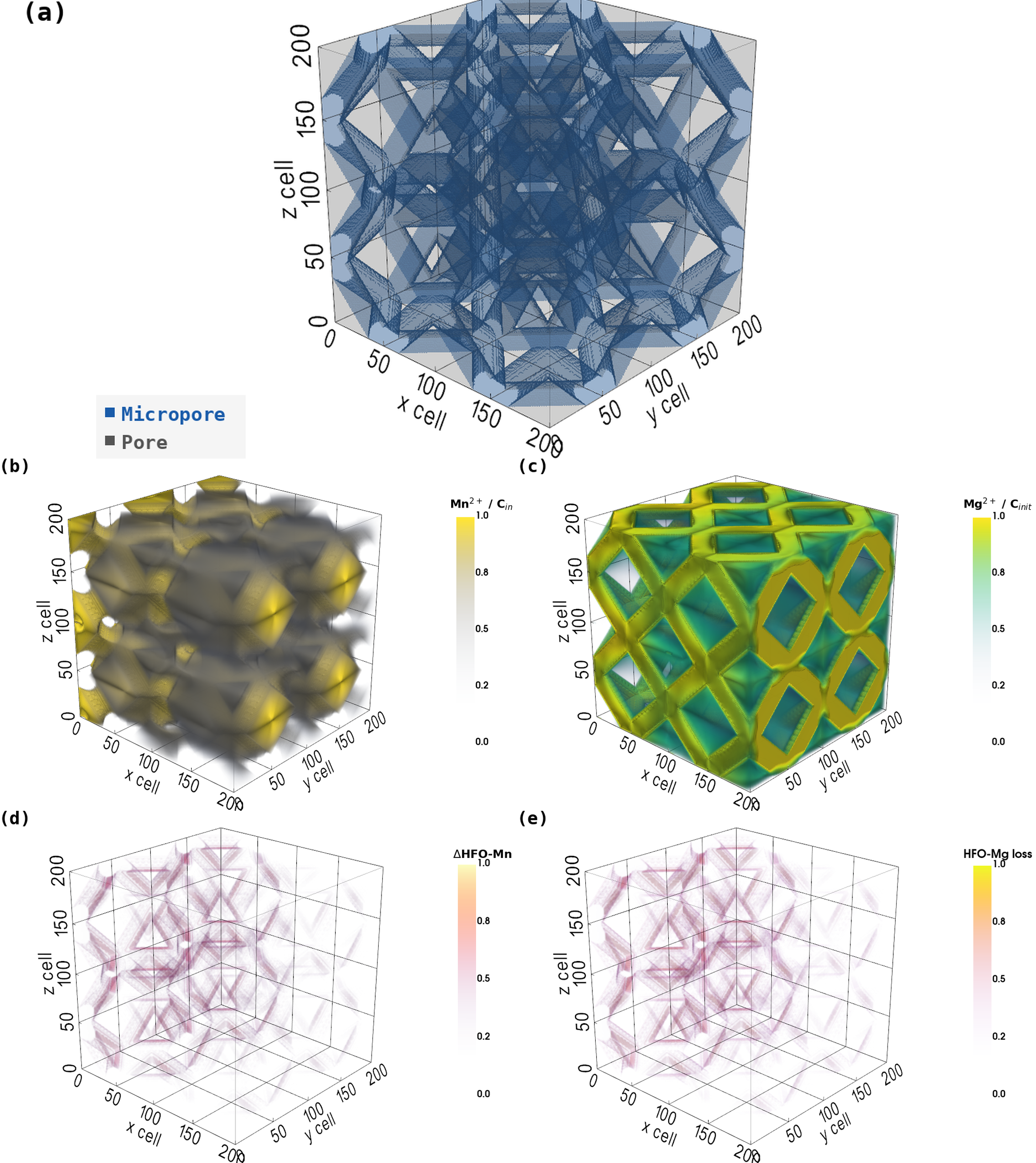}
    \caption{Competitive Mn$^{2+}$ and Mg$^{2+}$ sorption in the Kelvin case at
    $t = 120$~s.
    (a)~Segmented geometry: micropore diamond lattice in blue, pore space in grey.
    (b)~Normalised dissolved Mn$^{2+}$ / $C_{\mathrm{in}}$; the advancing front enters
    from the left.
    (c)~Normalised dissolved Mg$^{2+}$ / $C_{\mathrm{init}}$.
    (d)~Incremental adsorbed Mn, summed over strong and weak HFO sites,
    concentrated in the microporous diamond-lattice channels.
    (e)~Magnitude of the displaced adsorbed Mg, showing where Mn$^{2+}$ has
    replaced part of the initially sorbed Mg$^{2+}$ inventory on weak HFO sites.}
    \label{fig:3DKelvin}
\end{figure}

\begin{figure}[htbp]
    \centering
    \includegraphics[width=0.9\linewidth]{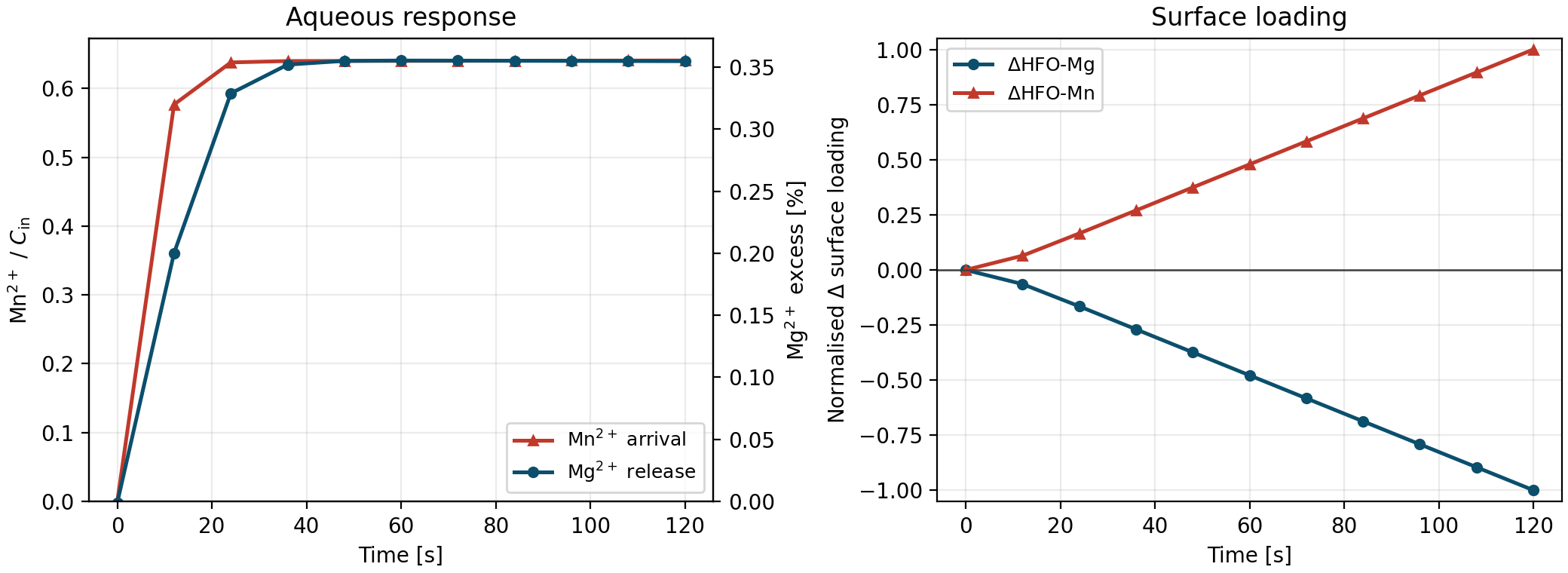}
    \caption{Temporal evolution of domain-mean fields for the Kelvin case
    ($t = 0$ to $120$~s, sampled every $12$~s).
    Left: Mn$^{2+}$ normalised by the inlet concentration and the Mg$^{2+}$
    excess above the background concentration, both pore-volume weighted.
    Right: normalised signed change in surface-area-weighted adsorbed amount
    for summed strong- plus weak-site HFO-Mn and weak-site HFO-Mg, defined relative to
    their values at $t = 0$.}
    \label{fig:KelvinTemporal}
\end{figure}

Figure~\ref{fig:3DKelvin} shows the geometry, volume renders of the two
competing dissolved species, and the adsorbed-amount changes at $t = 120$~s.
Figure~\ref{fig:KelvinTemporal} follows the pore-volume weighted aqueous
Mn$^{2+}$, the aqueous Mg$^{2+}$ released above the background concentration,
and the surface-area-weighted strong- plus weak-site HFO loadings over the
120~s simulation. The adsorbed-amount panel quantifies the competitive
exchange: HFO-Mn accumulates behind the front, while the HFO-Mg inventory
decreases wherever the Mn-bearing water has reached the sorbent.

\FloatBarrier
\clearpage
\subsection{\texorpdfstring{3D Estaillades Limestone: Sr$^{2+}$ in a Deep
Geological Repository}{3D Estaillades Limestone: Sr2+ in a Deep Geological
Repository}}

This case uses a three-phase segmented Estaillades limestone image
\citep{menke2021upscaling} with dimensions $500 \times 500 \times 500$ and
voxel size $4~\mu\mathrm{m}$, in which the phases are pore ($1$), micropore
($2$), and solid ($3$). Transport is solved at the native voxel resolution on
64 MPI ranks, which places the nonlinear surface exchange in a geologically
realistic carbonate pore structure.

The chemistry models Sr$^{2+}$ leachate from a waste package invading
Ca$^{2+}$-bearing limestone formation water. Before transport begins, the
surface sites are equilibrated locally with a $10^{-3}$~mol~L$^{-1}$ Ca$^{2+}$
solution at pH~7,
so the initial sorbed inventory is Ca-dominated. The two competing surface
reactions are:
\begin{align}
    \mathrm{Surf_a H} + \mathrm{Ca^{2+}} &\rightleftharpoons \mathrm{Surf_a Ca^{+}} + \mathrm{H^{+}}, \quad \log K = -5.85 \\
    \mathrm{Surf_a H} + \mathrm{Sr^{2+}} &\rightleftharpoons \mathrm{Surf_a Sr^{+}} + \mathrm{H^{+}}, \quad \log K = -5.5
\end{align}
The single \texttt{Surf\_a} site has a density of
$2.4\times10^{-6}$~mol~m$^{-2}$, and Sr$^{2+}$ binds slightly more strongly
than Ca$^{2+}$ ($\log K = -5.5$ against $-5.85$). The domain and the outlet
boundary start at Ca$^{2+} = 10^{-3}$~mol~L$^{-1}$ with no Sr$^{2+}$, and the
inlet carries Sr$^{2+} = 10^{-4}$~mol~L$^{-1}$ with no Ca$^{2+}$, both at pH~7.

This produces a chromatographic exchange pattern in which a Ca$^{2+}$
desorption wave precedes the Sr$^{2+}$ adsorption front, and Sr$^{2+}$ is
retarded less than a single-species calculation would give because Ca$^{2+}$
still competes for sites ahead of that front. Over the saved output sequence ($t = 0, 20, \ldots, 120$~s) the
porosity-weighted mean Sr$^{2+}$ concentration rises from zero to
$0.865\,C_{\mathrm{in}}$ while the Ca$^{2+}$ concentration falls from
$1.000$ to $0.095\,C_{\mathrm{init}}$. On the surface,
\mbox{$\Delta$\texttt{Surf\_aSr+}} increases monotonically and
\mbox{$\Delta$\texttt{Surf\_aCa+}} becomes progressively more negative,
quantifying the net replacement of Ca$^{2+}$ by Sr$^{2+}$.

\begin{figure}[htbp]
    \centering
    \includegraphics[width=0.85\linewidth]{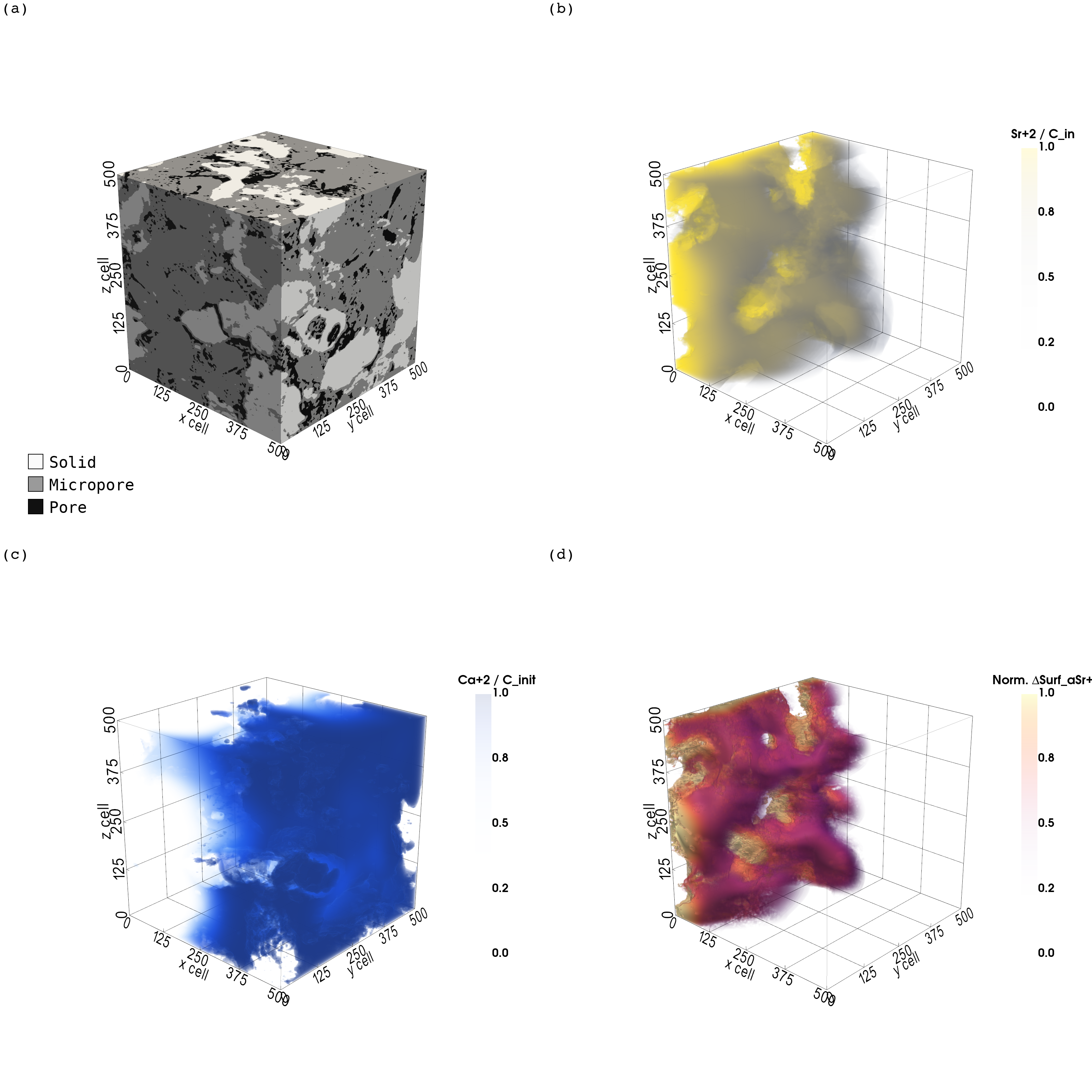}
    \caption{3D volume renders of the Estaillades DGR case at $t = 40$~s.
    (a)~Segmented geometry: solid in white, micropore in grey, and pore in black.
    (b)~Normalised dissolved Sr$^{2+}$ / $C_{\mathrm{in}}$ in the mobile-water
    domain; the plume advances from the inlet (left face).
    (c)~Normalised dissolved Ca$^{2+}$ / $C_{\mathrm{init}}$ in the same domain;
    the complementary depletion wave shows that Sr$^{2+}$ has displaced Ca$^{2+}$
    in the swept region.
    (d)~Incremental adsorbed Sr$^{2+}$ (\mbox{$\Delta$\texttt{Surf\_aSr+}}),
    relative to $t = 0$ and normalised by its maximum at $t = 40$~s. The
    distributed loading shows Sr$^{2+}$ accumulating throughout the reactive
    microporous phase.}
    \label{fig:3DEstaillades}
\end{figure}

\begin{figure}[htbp]
    \centering
    \includegraphics[width=0.9\linewidth]{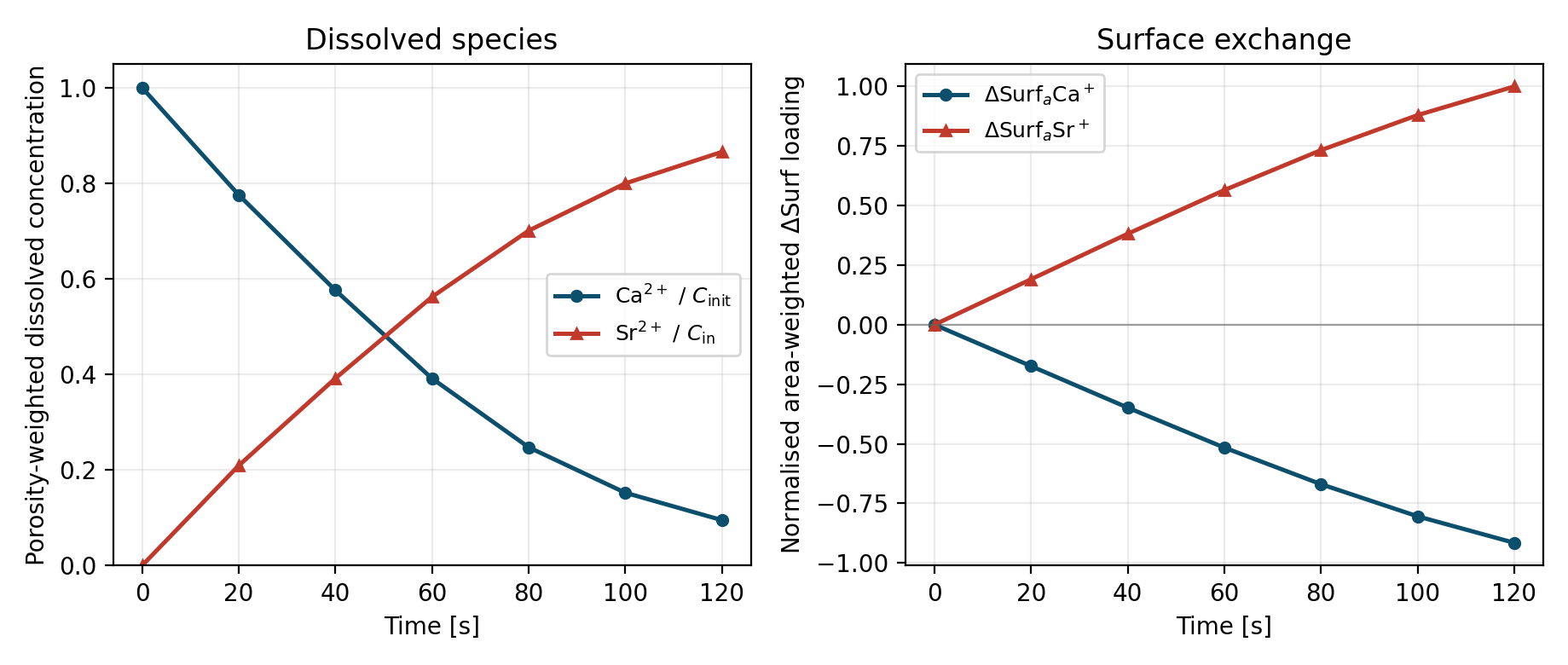}
    \caption{Temporal evolution of weighted fields for the Estaillades DGR
    case ($t = 0$ to $120$~s, write interval $= 20$~s).
    Left: porosity-weighted dissolved Ca$^{2+}$ and Sr$^{2+}$ normalised by
    $C_{\mathrm{init}}$ and $C_{\mathrm{in}}$, respectively.
    Right: normalised incremental surface-area-weighted adsorbed amount
    (\mbox{$\Delta$\texttt{Surf\_aCa+}} and \mbox{$\Delta$\texttt{Surf\_aSr+}}),
    defined relative to their values at $t = 0$ and divided by the largest
    absolute surface-loading change over the plotted sequence.
    The curves show the chromatographic exchange: Sr$^{2+}$ invades and
    accumulates on the surface while Ca$^{2+}$ is displaced from both the
    aqueous and sorbed phases.}
    \label{fig:EstailladesTemporal}
\end{figure}

Figure~\ref{fig:3DEstaillades} shows the geometry and 3D volume renders of the
dissolved and adsorbed fields at $t = 40$~s.
Figure~\ref{fig:EstailladesTemporal} summarises the weighted exchange history.

\FloatBarrier
\section{Discussion}

\subsection{Formulation Choices and Conservative Structure}

The central modelling choice in this work is to use one fixed porosity field,
$\theta$, in the finite-volume transport equation and in the PHREEQC cell
volumes alike, which keeps aqueous accumulation, transport, and equilibrium
chemistry on the same volumetric basis. The reactive surface is prescribed
independently through the explicit cell field $a_{\mathrm{surf}}$, so the
chemical storage volume and the amount of reactive surface remain separately
defined material properties.

This approach extends earlier micro-continuum reactive transport formulations
to the case where the reactive surface area is an unresolved cell property
rather than an explicitly meshed boundary. In existing pore-scale DBS solvers,
including \nolinkurl{porousMedia4Foam} and earlier GeoChemFoam solvers,
reactive uptake is typically imposed as a flux condition on the diffuse
fluid-solid interface. This requires the reactive surface to coincide with the resolved image voxel
boundary, which is a good approximation when the dominant reactive area lies at
the contact between macropore and grain. In multi-scale carbonate images the
microporous phase holds a large internal reactive area that lies inside the
coarse cell rather than on its boundary. The cell-based formulation carries
that area as an intrinsic volumetric density
$a_{\mathrm{surf}}^{\mathrm{intrinsic}}$, scaled by the local phase fraction
and passed to PHREEQC as a cell-level property.

The closed-batch balance isolates the round-trip transfer of aqueous and
adsorbed Ca between the finite-volume fields and PhreeqcRM: with transport and
boundary fluxes disabled, the combined inventory
$M=\theta C+a_{\mathrm{surf}}\Gamma$ should remain unchanged during local
equilibration. The transient 1D benchmark then tests the coupled transport
response, while its mesh- and timestep-refinement study quantifies the combined
spatial, temporal, and sequential-coupling errors.

\subsection{Geometric Scalability}

Because the validation ladder separates aqueous--surface state transfer,
coupled transport, surface-area geometry, and microporosity, the evidence can
be attributed to the layer that produced it. The closed-batch residual remains
below $10^{-9}$ in relative magnitude, the reported transient profiles differ
from the finest combined result by $3.01\%$ and $1.05\%$ for the
surface-enabled and aqueous-only cases, and the interface-area construction is
accurate to sub-percent levels against analytic spheres at
$R/\Delta x\ge16$ and against circular posts at all four tested resolutions. The
intrinsic micropore density enters only in the cases that use it.

The transition from 2D to 3D introduces qualitatively different transport paths
and surface-loading distributions. In the 2D case, open fractures provide
preferential pathways for dissolved Pb$^{2+}$, while strong-site Pb occupancy
varies across the surrounding microporous matrix and is enhanced near the inlet
and fracture intersections. In the 3D cases, the tortuosity of the
pore network produces a more dispersed front and a more heterogeneous
surface-loading pattern. This spatial heterogeneity in surface uptake controls the effective retardation
factor seen at the continuum scale, and following it requires a simulation that
resolves the local connectivity of the image.

The direct resolved-to-DBS comparison spans $Pe=0.1$ to $10$, with no
observed breakdown in agreement at the higher $Pe$ values tested. This
behaviour is consistent with the spatially uniform matrix closure and
repeated geometry of the benchmark, whose limited heterogeneity may help
preserve the accuracy of the two-scale representation as advection becomes
more dominant.

\subsection{Computational Cost Drivers}

PHREEQC is called once per active cell per timestep, with a full speciation
and surface-equilibration calculation for each cell. The chemistry overhead
therefore depends primarily on the number of active cells, the geochemical
system, and the frequency of the chemistry call. The resolved-wall and DBS
comparisons use the same nominal image resolution and quantify physical
agreement. A controlled performance comparison across resolutions is outside
the scope of this study. For larger grids or more complex geochemical systems,
masking near-solid cells
out of the PhreeqcRM mapping (here at
$\theta \le \theta_{\mathrm{mask}} = 10^{-3}$) is the main lever on that cost.
The threshold is held fixed across all examples here.

\subsection{Limitations and Future Work}

The intrinsic micropore end member $s_{\mathrm{micro}}$ is prescribed from
complementary measurements such as nitrogen BET adsorption, mercury intrusion
porosimetry, or finer-resolution image analysis \citep{noiriel2009changes}.
The segmented image determines the geometric fraction $f_{\mathrm{micro}}$,
and their product gives the bulk-volume cell contribution
$a_{\mathrm{surf}}^{\mathrm{intrinsic}}=f_{\mathrm{micro}}s_{\mathrm{micro}}$.
Uncertainty in $s_{\mathrm{micro}}$ is therefore a leading physical uncertainty
in microporous applications. This uncertainty is not specific to the
cell-based formulation: a resolved-wall representation constructed from the
same segmented image also cannot recover reactive area below the image
resolution without complementary characterisation. Linking image-derived
phase fractions to measured or computed phase-specific surface densities is
therefore required to constrain predicted uptake.

The upscaled DBS treatment carries unresolved matrix properties as local
cell-averaged quantities. When diffusive exchange is fast compared with the
advective residence time, the intrinsic surface area acts as a locally
accessible reactive reservoir. At larger $Pe$, advective bypass and finite
matrix-exchange times can restrict access to that area, so reactive uptake
depends on its time-dependent accessibility as well as its amount. This
transport-upscaling limitation is distinct from the assumption of local
chemical equilibrium within each active cell: local equilibration does not
ensure that solute reaches the unresolved reactive surface. Media with more
heterogeneous matrix properties, connectivity, or exchange length scales
therefore require their own $Pe$-resolved assessment.

The resolved-to-DBS comparison tests this practical consequence of the
intrinsic-area representation directly. For this micromodel, the small
inventory differences and modest axial-profile differences show that replacing
explicitly resolved matrix pore walls with cell-supported reactive area does
not materially alter domain-scale Li uptake across the tested $Pe$ range.

Transport and chemistry are coupled by a first-order Lie splitting, which
introduces a splitting error that scales with the timestep and the stiffness of
the reaction terms. For the surface-complexation systems used here, that error
sits within the benchmark agreement reported in Section~\ref{sec:validation},
but it may become significant for faster reactions or stiffer geochemical
systems. A second-order Strang splitting \citep{strang1968construction} is the
natural extension.

Restart depends on the serialised PHREEQC state written with each complete
checkpoint, because the finite-volume \texttt{Surf\_*} fields alone cannot
reconstruct that state. The production $Pe$ sweep restarted from these
checkpoints across several jobs, and moving a run between machines requires the
field data and its matching PHREEQC dump to travel together.

The cases here use a fixed surface area and equilibrium surface complexation,
and cover competitive multi-component chemistry and a Li-selective multi-ion
brine within that scope. Evolving reactive area, mineral dissolution and
precipitation, kinetic surface reactions, and the wider range of mineral
surfaces and aqueous compositions found in natural systems lie outside it.
Extending the model to them will require benchmarks that separate transport
error, surface-area evolution, and reaction-model uncertainty while retaining
cell-level mass-balance checks.
\section{Conclusions}

This work presents a conservative DBS formulation for PHREEQC-coupled reactive
transport with cell-based surface complexation in image-based porous media,
validated across analytical, image-based, and multi-scale porous-media cases.

The key contribution is a cell-based representation of unresolved reactive
surface area that sits inside the DBS micro-continuum framework. One fixed
porosity field serves aqueous accumulation, transport, and the PHREEQC cell
volumes alike, which keeps the aqueous transport conservative and the local
mass balance closed cell by cell. The unresolved micropore surface enters as an
additive intrinsic volumetric density, so reactive uptake can be represented in
image phases whose internal surface area lies inside the coarse cell rather
than on its boundary.

The 1D tests preserve the closed-batch Ca inventory to a relative residual
below $10^{-9}$ and show approximately first-order mesh and timestep
convergence, while the surface-area reconstruction recovers analytic interface
area to sub-percent accuracy for 3D spheres at $R/\Delta x\ge16$ and
quasi-2D circular posts at all four tested resolutions. In the Li micromodel, the
intrinsic-area representation keeps both aqueous and sorbed inventories within
$0.6\%$ of the resolved result over $Pe=0.1$--10, while the sorbed streamwise
profile differs by up to $4.75\%$. Across the 2D and 3D image-based cases, the
model resolves the dissolved and adsorbed fields that follow from the local
pore geometry and competition between species for surface sites.

Surface complexation can therefore be carried conservatively on cells whose
reactive area is unresolved, freeing the reactive surface from the resolved
mesh boundary in multi-scale image-based simulation. This matters for reactive
porous-media problems in which selectivity and surface accessibility control
performance, among them contaminant retention, radionuclide migration, and
critical-mineral recovery by sorbent-bearing porous media. Constraining the intrinsic surface-area input against measured internal area is
the step that would most improve predictive accuracy, and extending the
formulation to evolving reactive area and kinetic surface chemistry would open
the dissolution and precipitation problems that the equilibrium treatment
excludes.

\clearpage
\section*{Supplementary Information}

+\section{One-Dimensional Numerical Convergence}

The spatial and temporal sensitivity of the 1D retardation
benchmark was evaluated at $t=10$~s for both the surface-enabled and matched
aqueous-only cases. The mesh study used $N_x=40$, 80, 160, and 320 cells with
$\Delta t=0.015625$~s. The timestep study used $N_x=320$ with
$\Delta t=0.25$, 0.125, 0.0625, 0.03125, and 0.015625~s. All other physical and
chemical inputs were unchanged from the benchmark reported in the main text.

Table~\ref{tab:1dConvergence} reports the relative $E_1$ difference between
each concentration profile and the next finer result. For the mesh comparison,
the finer profile was conservatively averaged onto the coarser cells before
evaluation. The successive differences decrease monotonically for both cases.
The observed orders range from 0.83 to 0.96 for the surface-enabled mesh
sequence and from 0.95 to 0.96 for its timestep sequence; the corresponding
aqueous-only ranges are 0.76--0.99 and 0.94--1.13. These values are consistent
with the first-order Euler time integration, upwind advection, and sequential
transport--equilibrium coupling used here. Because these are pairwise numerical
differences, they do not depend on a semi-infinite reference profile.

\begin{table}[htbp]
    \centering
    \caption{Mesh and timestep refinement for the 1D retardation benchmark at
    $t=10$~s. Values are relative $E_1$ differences (\%) from the next finer
    result. A dash identifies the finest result in each sequence.}
    \label{tab:1dConvergence}
    \begin{tabular}{llrrr}
        \toprule
        Study & Resolution & $N_x$ & Surface-enabled & Aqueous-only \\
        \midrule
        Mesh & $\Delta x=1.000~\mu$m & 40  & 0.949 & 0.583 \\
             & $\Delta x=0.500~\mu$m & 80  & 0.532 & 0.344 \\
             & $\Delta x=0.250~\mu$m & 160 & 0.274 & 0.173 \\
             & $\Delta x=0.125~\mu$m & 320 & --    & --    \\
        \midrule
        Timestep & $\Delta t=0.250000$~s & 320 & 1.444 & 0.465 \\
                 & $\Delta t=0.125000$~s & 320 & 0.747 & 0.213 \\
                 & $\Delta t=0.062500$~s & 320 & 0.383 & 0.109 \\
                 & $\Delta t=0.031250$~s & 320 & 0.198 & 0.057 \\
                 & $\Delta t=0.015625$~s & 320 & --    & --    \\
        \bottomrule
    \end{tabular}
\end{table}

The $80$-cell, $\Delta t=0.25$~s discretisation used for the main validation
differs from the finest combined result
($N_x=320$, $\Delta t=0.015625$~s) by $3.01\%$ and $1.05\%$ for the
surface-enabled and aqueous-only profiles, respectively. The monotonic
pairwise differences and observed orders quantify the combined spatial,
temporal, and sequential-coupling errors in the reported discretisation.

\section{Reactive-Area Construction and Capacity Mapping}

For each ordered pair of segmented labels, the owner and other phase define
a pair mask through their union. In \texttt{occupancyGaussian}, the
Gaussian-filtered owner indicator is divided by the Gaussian-filtered pair
mask to form a normalised occupancy field. The $0.5$ level is extracted
within the pair mask.

In three dimensions, reconstructed triangle area is assigned to computational
cells and divided by bulk cell volume. Paper-local three-dimensional
deposition probes owner points at offsets from $0.2$ to $3.0$ voxels along
the triangle normal, followed by a nearest-owner search in the surrounding
voxels. If this search finds no owner voxel, the triangle centroid supplies
the mapping point.

In quasi-two-dimensional images, reconstructed contour length is multiplied
by the physical out-of-plane thickness before division by bulk cell volume.
The serial construction probes offsets from $0.2$ to $1.5$ voxels along the
contour normal and uses the contour-piece midpoint if the probes find no
owner point. Weighted redistribution assigns each contour piece to
owner-side first-shell contact cells, with weights based on distance and
contact support. Candidate weights are normalised for each piece.
Local quasi-two-dimensional construction uses contact-support-based mapping
and deposits the weighted shares assigned to rank-owned cells. Pieces with
empty support contribute zero area to the deposited field. Both dimensional
routes return m$^2$\,m$^{-3}_{\mathrm{bulk}}$. The field is constructed before
flow and held fixed throughout reactive transport.

For intrinsic area, $s_k$ is defined per unit volume of phase $k$
(m$^2$\,m$^{-3}_{k}$). The geometric phase fraction $f_k$ converts this
end-member density to a bulk-volume cell contribution:
$a_{\mathrm{surf}}^{\mathrm{intrinsic}}=\sum_k f_k s_k$.
A mass-specific measurement is converted to phase-volume density using the
corresponding mass per unit phase volume. A density already expressed per
bulk cell volume represents the cell contribution directly.

The \texttt{phreeqcMixture} builds used for the reported calculations supply
PhreeqcRM \citep{parkhurst2015phreeqcrm} with area per representative bulk
litre, obtained by dividing $a_{\mathrm{surf}}$ in
m$^2$\,m$^{-3}_{\mathrm{bulk}}$ by $10^3$. For surface master $k$, the site
density $N_k$ (mol\,m$^{-2}$) gives the surface-master inventory per bulk
volume as $N_k a_{\mathrm{surf}}$. These bulk-volume conversions apply
without division by $\theta$. The inventory units are
\[
\left[\frac{\mathrm{m}^2}{\mathrm{m}^3_{\mathrm{bulk}}}\right]
\left[\frac{\mathrm{mol}}{\mathrm{m}^2}\right]
=\left[\frac{\mathrm{mol}}{\mathrm{m}^3_{\mathrm{bulk}}}\right].
\]

The chemistry mask threshold is set by \texttt{epsChemistryMin} in
\texttt{constant/thermoPhysicalProperties}, and every calculation reported here
uses $\theta_{\mathrm{mask}}=10^{-3}$.

\section{Discrete PhreeqcRM Coupling}

The surface-complexation cases select \texttt{mixtureType phreeqcMixture} in
\texttt{constant/thermoPhysicalProperties}. The active mapping between
finite-volume cells and PhreeqcRM cells is established once and does not change
during the run. Reconstruction is needed only when a threshold crossing changes
the active-cell map or its topology, or when the fixed area capacity changes.

Each OpenFOAM process constructs PhreeqcRM with one worker thread. Domain
decomposition provides process-level parallelism, and \texttt{runCells()}
advances only the chemistry cells held by that rank. Each case is decomposed
according to the \texttt{NPX}, \texttt{NPY}, and \texttt{NPZ} entries in its
\texttt{system} directory. Every case uses its case-local
\texttt{constant/GeoChem.dat} database. The image-based cases use kinematic
viscosity $\nu = 10^{-6}$~m$^2$\,s$^{-1}$ and kinematic pressure drop
$\Delta p = 5\times10^{-5}$~m$^2$\,s$^{-2}$, and write output every 10~s for the 1D
benchmark, 2000~s for the fracture case, and 20~s for the Estaillades case.

\subsection{Checkpoint and Restart Lifecycle}

At an output time, a checkpoint-enabled run first invalidates the old
completion marker collectively, writes a new rank-local dump, writes the
finite-volume fields, and publishes a completion marker only after the field
write succeeds. A crash before publication therefore leaves the partially
updated fields and chemistry state without a marker, and restoration rejects
them. The PhreeqcRM dump and marker must be retained with the corresponding
finite-volume time directory; the \texttt{Surf\_*} fields alone do not
reconstruct the complete internal state.

\section{Short Two-Scale Reconstruction Sensitivity}

The two-scale calculation is a matched short-interval sensitivity test between
two DBS cases that differ only in the surface field. Both use a
$3600\times1200\times1$ grid with $5~\mu$m cells,
16 MPI ranks, $Pe=1$, $D=10^{-9}$~m$^2$\,s$^{-1}$, and prescribed Darcy velocity
$5.46297\times10^{-5}$~m\,s$^{-1}$. The surface field is the only input that
differs, and the other 495 input files were verified as identical.

Both short cases use their case-local \texttt{GeoChem.dat} together with the
same illustrative Li-selective sorbent chemistry. The surface reactions are
$\mathrm{Sor\_aH}+\mathrm{Li^+}\rightleftharpoons\mathrm{Sor\_aLi}+\mathrm{H^+}$
($\log K=-5.8$), and the corresponding Na$^+$, K$^+$, and Mg$^{2+}$ reactions
have $\log K=-7.2$, $-7.0$, and $-7.4$, respectively. Initially Li-free brine
contains 0.10~M Na, 0.01~M K, 0.01~M Mg, $10^{-12}$~M Li, and 0.13~M Cl at pH
7; the inflow changes Li to $10^{-3}$~M and Cl to 0.131~M. These parameters
apply to the short $Pe=1$ sensitivity test only.

Three common checkpoints at 0.0058517566, 0.11703513, and 0.99564208~s meet
the predefined material-response limits. The maximum absolute differences
are $0.000647\%$ in integrated area, $0.000703\%$ in aqueous Li inventory,
$0.2110\%$ in sorbed Li inventory, $0.00172\%$ in the aqueous profile-shape
diagnostic, $0.5147\%$ in the sorbed profile-shape diagnostic, and $0.03993\%$
in fixed-reference-area sorbed-response L2. The two profile-shape diagnostics
compare independently normalised profiles and are not the relative $E_1$
measure used for validation. Shifts in the aqueous and sorbed cumulative-inventory median positions
($x_{50}$) are zero. The native-cell
sorbed inventory-density L2 is $3.61$ to $4.49\%$, dominated by the deliberate
relocation of area, so the agreement is in integrated response rather than
pointwise. The last accepted checkpoint covers $0.489\%$ of the configured
$203.53$~s horizon, so the test establishes early material-response agreement.

\section{Reproducibility and Provenance Entry Points}

The paper directory contains self-contained case definitions and source-data
references. From the repository root, the provenance verifier is run with
\begin{verbatim}
./validate_accepted_nonkelvin_artifacts.sh
\end{verbatim}
This command runs the verifier tests and validates
\texttt{figures/accepted\_nonkelvin\_provenance.json}. The manifest binds the
reported 1D, fracture, and Estaillades figures to their source data, renderers,
sidecars, and evidence records by path, size, and SHA-256.

The case-level entry point for the 1D benchmark is
\texttt{runPhase8AnalyticalBenchmark.sh} in
\nolinkurl{cases/1d_linear_retardation}. The mesh and timestep sequences are
reproduced by \texttt{runConvergenceStudy.sh}; the resulting metrics and exact
runtime identities are retained in \texttt{convergence\_results.json} and
\texttt{convergence\_provenance.json}. The fracture and Estaillades cases are
run through the \texttt{runAll.sh} scripts in
\nolinkurl{cases/2d_4StraightFractures} and \nolinkurl{cases/3d_estaillades},
each of which also provides separate mesh, flow, reactive-transport,
processing, and cleanup scripts. The short
two-scale record is in the \texttt{results} directory of
\nolinkurl{cases/two_scale_micromodel_sorption_validation}; it records exact
checkpoints, thresholds, collection hashes, runtime identity, and analyser
identity. The paper sources and both PDFs are built with \texttt{build.sh}. The
flat Springer submission archive is generated transactionally with
\texttt{export\_submission.sh OUTPUT.zip}; its manifest and SHA-256 inventory
are included in the archive.

\backmatter

\section*{Statements and Declarations}

\textbf{Funding:} This work was supported by the Natural Environment Research
Council (NERC) and Nuclear Waste Services (NWS) through the GeoSAFE grant
(reference NE/Y002504/1). Opinions, findings, conclusions or recommendations
expressed are those of the authors and do not necessarily reflect the views
of NERC or NWS.

\textbf{Competing interests:} The authors have no relevant financial or
non-financial interests to disclose.

\textbf{Ethics approval:} Not applicable.

\textbf{Consent to participate:} Not applicable.

\textbf{Consent for publication:} Not applicable.

\textbf{Availability of data and materials:} The data and materials supporting
this study are available from the corresponding author on reasonable request.

\textbf{Code availability:} The GeoChemFoam 5.2 codebase is publicly available
at \url{https://github.com/GeoChemFoam/GeoChemFoam-5.2}. The solver modifications,
paper-specific cases, simulation scripts, and post-processing code are available
from the corresponding author on reasonable request.

\textbf{Authors' contributions:} \textbf{Hannah P. Menke:} Conceptualisation,
Methodology, Software, Investigation, Formal analysis, Writing -- original
draft, Funding acquisition. \textbf{Julien Maes:} Conceptualisation,
Methodology, Writing -- review \& editing, Funding acquisition.
\textbf{Florian Doster:} Conceptualisation, Writing -- review \& editing,
Funding acquisition.

\end{document}